\documentclass[pra,twocolumn,superscriptaddress,showpacs,amsmath,amstex,amssymb,citeautoscript]{revtex4-1}

\usepackage[T1]{fontenc}
\usepackage{booktabs}
\usepackage[utf8]{inputenc}
\usepackage{lipsum}
\usepackage{amsmath}
\usepackage{amssymb}
\usepackage{bbm}
\usepackage{braket}
\usepackage{xcolor}
\usepackage{pifont}
\usepackage[mathscr]{euscript}
\usepackage[shortlabels]{enumitem}
\usepackage{graphicx}
\usepackage{lipsum}
\allowdisplaybreaks
\usepackage{float}
\usepackage{graphicx}
\usepackage{dsfont}
\usepackage{comment}
\usepackage[colorlinks=true]{hyperref}  
\hypersetup{
    bookmarks=true,         
    unicode=false,          
    pdftoolbar=true,        
    pdfmenubar=true,        
    pdffitwindow=false,     
    pdfstartview={FitH},    
    pdftitle={TITLE},    
    pdfauthor={Someone et al.},     
    pdfsubject={},   
    pdfcreator={},   
    pdfproducer={}, 
    pdfkeywords={} {} {}, 
    pdfnewwindow=true,      
    colorlinks=true,       
    linkcolor=blue, 
    citecolor=blue,        
    filecolor=magenta,      
    urlcolor=blue           
}

\newcommand{\equref}[1]{Eq.~(\ref{#1})}
\newcommand{\equsref}[2]{Eqs.~(\ref{#1}) and (\ref{#2})}

\newcommand{\figref}[1]{Fig.~\ref{#1}}
\newcommand{\refcite}[1]{Ref.~\onlinecite{#1}}

\renewcommand{\approx}{\simeq}

\definecolor{wrongultramarine}{rgb}{1,0.5,0}

\renewcommand{\vec}[1]{\boldsymbol{#1}}

\begin{document}

\title{Altermagnetic spin textures: \\ Emergent electrodynamics, quantum geometry, and probes}

\author{Constantin Schrade}
\affiliation{Hearne Institute of Theoretical Physics, Department of Physics \& Astronomy, Louisiana State University, Baton Rouge LA 70803, USA}

\author{Mathias S.~Scheurer}
\affiliation{Institute for Theoretical Physics III, University of Stuttgart, 70550 Stuttgart, Germany}

\begin{abstract}
Emergent electrodynamics arising from spatially and temporally varying magnetic textures provides a framework for spin control in quantum materials. While this principle is established for ferromagnetic and antiferromagnetic textures, its
consequences for altermagnets---magnetic orders with vanishing net magnetization but finite spin splitting---remain largely unexplored. In this work, we develop an effective low-energy theory of itinerant electrons coupled to smoothly varying altermagnetic spin textures. In the adiabatic regime, we show that altermagnetic textures generate additional emergent electromagnetic fields and quantum-geometric effects that are absent in conventional magnetic systems. These effects include emergent Zeeman fields that encode the structure of the altermagnetic order parameter, enabling local spin manipulation and a way to distinguish different altermagnetic orders. Moreover, we demonstrate a quantum-metric-induced, spin-dependent electron lensing effect that provides a mechanism for spin filtering, and discuss the local admixture of effective odd-parity magnetic components. Our results suggest that textured altermagnets could serve as a versatile resource for spintronics functionalities and a probe of altermagnetism. 
\end{abstract}

\maketitle

\section{Introduction}
Spin textures play a central role in condensed matter systems, ranging from magnetic textures in spintronics~\cite{parkin2008magnetic,fert2013skyrmions} to moiré lattices with spatially varying effective spin degrees of freedom~\cite{PhysRevLett.132.096602,PhysRevB.110.035130,PhysRevB.110.115114,PhysRevB.111.125122,guerci2026}. A particularly interesting feature of such textures is the emergence of effective electrodynamics for itinerant electrons, which has been studied for ferromagnets~\cite{Volovik1987,PhysRevLett.93.096806,PhysRevLett.102.067201,nam2009,PhysRevLett.98.246601,schulz2012,qxnw-8q4y} and antiferromagnets~\cite{PhysRevB.86.245118,PhysRevB.91.144421,okabayashi2015theory,PhysRevB.93.180408}. Intuitively, as an electron moves through a smoothly varying texture (with a characteristic length scale much larger than the Fermi wavelength), its spin must continuously align with the effective magnetic field, which generates additional forces and modifies its dynamics. These additional contributions can be mapped to emergent electromagnetic fields and a distortion of the underlying space in which the electrons propagate. From a lattice perspective, the same physics appears as a texture-dependent quenching of hopping amplitudes set by the relative orientation of neighboring spins. Emergent electrodynamics and quantum geometry of this kind, as well as current-induced torques~\cite{RevModPhys.91.035004,PhysRevLett.93.127204,PhysRevLett.100.226602,PhysRevB.83.054428,PhysRevLett.106.107206,PhysRevB.94.054409}, provide a powerful route to controlling spin degrees of freedom for 
spintronics applications~\cite{RevModPhys.76.323,RevModPhys.90.015005,jungwirth2025}.
\begin{figure}[!t]
    \centering
    \includegraphics[width=1\linewidth]{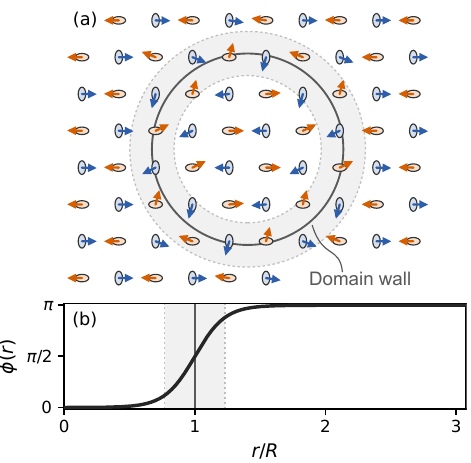}
\caption{
\textbf{Circular Néel domain wall separating two altermagnetic domains.}
(a) Real-space schematic of the Néel vector field, $\vec{n}(\boldsymbol{r})$, on the two sublattices 
($\tau_z=\pm$; orange and blue). 
Arrows indicate the local in-plane spin orientation. 
The shaded annulus denotes the domain wall. 
The solid circle marks the domain wall radius, $R$.
(b) Radial rotation profile, $\phi(r)$, describing the 
$\pi$-rotation of $\vec{n}$ across the wall. 
The wall center is at $r = R$.
The illustrations are schematic and not to scale. In the paper, we assume a texture with $a \ll w \ll R$ ($a$ is the lattice constant and $w$ is the wall width).
}
    \label{fig:1}
\end{figure}

Another exciting opportunity in spintronics that emerged more recently are altermagnets~\cite{PhysRevX.12.040501,Review3,magnetism5030017,PhysRevX.12.040002,RafaelReview}. In the simplest collinear cases, these magnetically ordered states are characterized by two magnetic sublattices which are related by symmetries that are---as opposed to antiferromagnets---not translation or inversion. A direct consequence is a finite spin splitting of the electronic bands while retaining vanishing net magnetization, which makes altermagnets attractive for spintronics applications~\cite{jungwirth2025}. Although spatially (e.g., domain walls) and possibly also temporally varying textures of the altermagnetic order parameter, $\vec{n}(\vec{r},t)$, are expected to be ubiquitous in real samples, in line with recent experiments~\cite{dp7v-qszq,NanoscaleImaging}, there are only very few theoretical works studying them~\cite{DomainWallsSinova, PhysRevLett.133.196701,PhysRevLett.134.176401,PhysRevB.111.064422}; these studies focus on the dynamics and form of the magnetic order parameter fields themselves, rather than on the impact on the electronic spectrum. Thus far, electronic properties of inhomogeneous altermagnets have primarily been studied in the context of atomic-scale defects~\cite{PhysRevB.111.174436,f6nc-vsnx,PhysRevB.110.205114,PhysRevB.111.035132,PhysRevB.110.165413,PhysRevB.111.L100502,PhysRevB.108.054510,2025arXiv251019943S}, as well as in studies on strain-induced responses~\cite{radhakrishnan2026,venderbos2025}.

In this work, we fill this gap and develop a theory for the impact of smoothly varying altermagnetic order parameters on the \textit{electronic} properties. This is particularly relevant for altermagnets where the impact of the magnetic moments on the electronic bands are non-trivial and their key defining feature. We will contrast the emergent electromagnetic fields and curved spacetime with the ferromagnetic and antiferromagnetic cases, revealing that there are novel contributions unique to altermagnets. We show that this not only allows to identify altermagnetic order but also to distinguish between different forms ($d$- versus $g$-wave) of altermagnetic order parameters. While our theory applies to generic textures, we primarily illustrate our findings using a circular domain wall, as illustrated in \figref{fig:1}.

\section{Minimal model}
As a starting point, we use the following minimal model for an altermagnetic spin texture, 
\begin{equation}
 \hat{\mathcal{H}}=\varepsilon_{0,\hat{\boldsymbol{p}}} \tau_0 +{t}_{x,\hat{\boldsymbol{p}}} \tau_x + {t}_{z,\hat{\boldsymbol{p}}} \tau_z  + J \tau_z \,\vec{n}(\boldsymbol{r},t) \cdot \vec{s}, 
\label{Eq1}
\end{equation}
motivated by earlier works on uniform altermagnets~\cite{PhysRevB.110.144412,PhysRevB.111.174436,839n-rckn}.
Here, the Pauli matrices $(\tau_{x},\tau_{y},\tau_{z})$ and $\vec{s}=(s_{x},s_{y},s_{z})^T$ act in sublattice and spin-space, respectively. Moreover, $\varepsilon_{0,\hat{\boldsymbol{p}}}$ describes
the sublattice-independent contribution to the Hamiltonian with $\hat{\boldsymbol{p}} = -i \hbar (\partial_x,\partial_y)^T$ being the momentum operator, whereas ${t}_{z,\hat{\boldsymbol{p}}}$ and ${t}_{x,\hat{\boldsymbol{p}}}$ are 
sublattice asymmetric and intersublattice contributions. For an altermagnet, ${t}_{z,\hat{\boldsymbol{p}}}$ transforms 
as a non-trivial, one-dimensional irreducible representation of the crystallographic point group. In the last term, $\vec{n}(\boldsymbol{r},t)$
describes the spatially ($\vec{r}=(x,y)^T$) and temporally ($t$) varying order parameter. When $\vec{n}(\boldsymbol{r},t)$ is spatially uniform and time-independent, \equref{Eq1} is readily diagonalized and yields the characteristic spin splitting of the bands with angular dependence determined by the functional form of ${t}_{z,\vec{p}}$ (as will become more explicit below). In particular, the splitting will vanish for $t_{z,\vec{p}} = 0$, which corresponds to antiferromagnetism.

To describe the effects of the altermagnetic spin texture, it is useful to move to a frame in which the quantization axis follows the local orientation of the Néel vector. We therefore parametrize the Néel vector as 
$
\vec{n}(\boldsymbol{r},t)
=
(
\sin\theta(\boldsymbol{r},t)\cos\phi(\boldsymbol{r},t),
\sin\theta(\boldsymbol{r},t)\sin\phi(\boldsymbol{r},t),
\cos\theta(\boldsymbol{r},t)
)^T
$ and define the unitary transformation
$
U(\boldsymbol{r},t)
=
\exp[
\frac{i}{2}\,\theta(\boldsymbol{r},t)
\left(
\sin\phi(\boldsymbol{r},t)s_x-\cos\phi(\boldsymbol{r},t)s_y
\right)
].
$
Under this transformation, the last term in Eq.\,\eqref{Eq1} becomes uniform since  
$
U^\dagger(\boldsymbol r,t)\,
(\vec{n}(\boldsymbol{r},t)\cdot\vec{s})\,
U(\boldsymbol{r},t)
=
s_z$. The information on the spatial and temporal variation of the altermagnetic spin texture are 
now contained in spin-dependent gauge fields, $\vec{A}(\boldsymbol{r},t)$, that shift the momentum operators as, 
$
\hat{\boldsymbol{\pi}} \equiv U^\dagger(\boldsymbol{r},t) \hat{\boldsymbol{p}} U(\boldsymbol{r},t)
= 
\hat{\boldsymbol{p}}
-
\vec{A}(\boldsymbol{r},t)
$
with
$
\vec{A}(\boldsymbol{r},t)
\equiv
-U^\dagger(\boldsymbol{r},t) [\hat{\boldsymbol{p}} U(\boldsymbol{r},t)]$.
The time-dependence of the order parameter further generates a temporal gauge field,
$A_0(\boldsymbol{r},t)\equiv i\hbar\,U^\dagger(\boldsymbol{r},t)[\partial_t U(\boldsymbol{r},t)]$. 
The rotating frame Hamiltonian, 
$
\hat{\mathcal{H}}_{\text{rot}}
=
U^\dagger(\boldsymbol{r},t) \hat{\mathcal H}(t) U(\boldsymbol{r},t)
-A_0(\boldsymbol{r},t)
$,
is then given by
\begin{equation}
 \hat{\mathcal{H}}_{\text{rot}}=\varepsilon_{0,\hat{\boldsymbol{\pi}}} \tau_0 +{t}_{x,\hat{\boldsymbol{\pi}}} \tau_x + {t}_{z,\hat{\boldsymbol{\pi}}} \tau_z  + J \tau_z s_{z} -A_0.
\label{Eq2}
\end{equation}
Here, we emphasize that the shifted momenta still commute, $[\hat{\pi}_i,\hat{\pi}_j] = U^\dagger(\boldsymbol{r},t)[\hat{p}_i,\hat{p}_j]U(\boldsymbol{r},t) = 0$. However, as we will see in the next section, this commuting property will be violated upon projecting the rotating frame Hamiltonian onto its low-energy subspace, leading to emergent electromagnetic fields.

Finally, it is instructive to introduce two examples that will guide the rest of this work. The first example is that of a $d$-wave altermagnet. Only keeping the leading power in momentum for every term, we set $\varepsilon_{0,{\boldsymbol{p}}}=\varrho_{0}({p}_x^2+{p}_y^2)$, 
${t}_{x,{\boldsymbol{p}}} = t_{x}$, and
${t}_{z,\boldsymbol{p}} = \varrho_{z}({p}_x^2-{p}_y^2)$. In the limit of a spatially uniform and time-independent $\vec{n}(\boldsymbol{r},t)$, 
this choice gives the usual dispersion of a conventional $d$-wave altermagnet. The second example is that of a $g$-wave altermagnet, where $\varepsilon_{0,\vec{p}}$ and ${t}_{x,\boldsymbol{p}}$ are chosen as before. Meanwhile, for ${t}_{z,\boldsymbol{p}}$ we choose a $g$-wave form, $t_{z,\boldsymbol{p}}=\bar{\varrho}_z p_x p_y(p^2_x-p^2_y)$.
At the operator level, we will use the representation 
$
\hat{t}_{z,\hat{\boldsymbol{p}}}
= 
\bar{\varrho}_z
\mathcal W[\hat{p}_x \hat{p}_y(\hat{p}^2_x-\hat{p}^2_y)]
$
where we have introduced the fully-symmetrized ``Weyl'' ordering, 
$\mathcal{W}(\hat O_1 \hat O_2 \hat O_3 \hat O_4 )
=\sum_{\pi\in S_4}
(\hat O_{\pi(1)}\hat O_{\pi(2)}\hat O_{\pi(3)}\hat O_{\pi(4)})/4!$
with $S_4$ being the symmetric group on four elements.
At this point, this representation is redundant, since both the momenta and shifted momenta commute. 
However, it will be useful later when discussing the semiclassical limit.

\section{Emergent electrodynamics}
Emergent electromagnetic fields are known to arise in ferromagnetic \cite{Volovik1987,PhysRevLett.93.096806,PhysRevLett.102.067201,nam2009,PhysRevLett.98.246601,schulz2012,qxnw-8q4y} and antiferromagnetic \cite{PhysRevB.86.245118,PhysRevB.91.144421,okabayashi2015theory,PhysRevB.93.180408} spin textures. Here, we show that altermagnets support qualitatively new emergent fields. These fields provide ways to distinguish (i) altermagnets from ferro- and antiferromagnets, and (ii) different altermagnetic order parameters. 

As a first step to derive the emergent electromagnetic fields, it will be useful to decompose the spin-dependent gauge fields into its longitudinal and transversal components, 
$
A_\mu(\boldsymbol{r},t)=A_\mu^\parallel(\boldsymbol{r},t)+A_\mu^\perp(\boldsymbol{r},t)
$
with
$
A_\mu^\parallel(\boldsymbol{r},t) \propto s_z
$,
$
A_\mu^\perp(\boldsymbol{r},t) \propto s_{x,y}
$,
and
$\mu\in\{0,x,y\}$. This decomposition allows us introduce momentum operators shifted by the longitudinal gauge field, 
$
\hat\Pi_i=\hat{p}_i-A_i^\parallel(\boldsymbol{r},t)
$
so that 
$
\hat{\pi}_i=\hat\Pi_i-A_i^\perp(\boldsymbol{r},t).
$
Also, we introduce $\hat{\Pi}_0=i\hbar\partial_t - A_0^\parallel(\boldsymbol{r},t)$.
One important feature of these newly defined momentum operators is that they do not generally commute.
Instead, we find that
$[\hat\Pi_x,\hat\Pi_y]=i\hbar \mathcal{B}(\boldsymbol{r},t)$ and $[\hat\Pi_0,\hat\Pi_i]=i\hbar \mathcal E_i(\boldsymbol{r},t)$  with the effective spin-dependent orbital (out-of-plane) magnetic fields and (in-plane) electric field components, 
\begin{equation}
\begin{split}
 \mathcal{B}(\boldsymbol{r},t)&= -s_{z}\,\frac{\hbar}{2}\, \vec{n}(\boldsymbol{r},t)\cdot[\partial_x\vec{n}(\boldsymbol{r},t)\times\partial_y\vec{n}(\boldsymbol{r},t)],\\
  \mathcal{E}_i(\boldsymbol{r},t)&= -s_{z}\,\frac{\hbar}{2}\, \vec{n}(\boldsymbol{r},t)\cdot[\partial_t\vec{n}(\boldsymbol{r},t)\times\partial_i\vec{n}(\boldsymbol{r},t)].
\end{split} \label{EmergentEMFields}
\end{equation}

We will now show that the itinerant electrons in our altermagnetic spin texture will effectively be subject to these $\mathcal{B}(\boldsymbol{r},t)$ and $\mathcal{E}_i(\boldsymbol{r},t)$ fields as well to \textit{an emergent Zeeman field}. For this purpose, we will assume that $J$ is a sufficiently large energy scale so that the low-energy states of our altermagnetic spin texture lie in the subspace where $S\equiv\tau_z s_z$ has eigenvalue $1$, and define the associated projection operator $P_{-}=(1-S)/2$. We will next study our rotating frame Hamiltonian in \equref{Eq2} upon being projected onto the $S=-1$ while, for now, neglecting virtual transitions to the high-energy $S=+1$ subspace.

We initially focus on our first example of the $d$-wave altermagnet. In this case, we find that the effective Hamiltonian (to $0^{\text{th}}$-order in $1/J$), $\hat{\mathcal{H}}^{(0)}_{\text{rot}}=P_{-}\hat{\mathcal{H}}_{\text{rot}}P_{-}$, evaluates to 
\begin{equation}
\hat{\mathcal{H}}^{(0)}_{\text{rot}}
=
\varepsilon_{0,\hat{\boldsymbol{\Pi}}} \sigma_0 
+
{t}_{z,\hat{\boldsymbol{\Pi}}} \sigma_z
+
\varrho_{0} V_0 \sigma_0
+
\varrho_{z} V_z \sigma_z
- \sigma_z a_0^\parallel. 
\end{equation}
Here, we already expressed the Hamiltonian as an effective $2$-band model using the Pauli matrices $\sigma_{j}$ acting on the low-energy degrees of freedom with basis states $(\tau=+,s=\downarrow)$ and $(\tau=-,s=\uparrow)$. In this expression, $s_z$ in $\hat{\vec{\Pi}}$ is implicitly replaced by $-\sigma_z$.
Furthermore, $
a_0(\boldsymbol{r},t) = (\hbar/2) [\hat{\mathbf z}\cdot\bigl(\vec n\times\partial_t\vec n\bigr)] / (1+n_z)$ is the temporal component of the projected longitudinal gauge field. Moreover, we have defined the scalar potentials
\begin{equation}
V_{0}
=
\hbar^2
\delta^{ij}g_{ij}(\boldsymbol{r},t), 
\quad
V_{z}
=
\hbar^2
\eta^{ij}g_{ij}(\boldsymbol{r},t).
\label{VzFordwave}\end{equation}
Here, $\eta^{ij}=\text{diag}(1,-1)_{ij}
=
\frac{1}{2!}
\frac{1}{\varrho_z}
\left.
\frac{\partial^2 t_{z,\boldsymbol{p}}}
{\partial{p}_i
\,
\partial{p}_j}
\right|_{\boldsymbol{p}=0}$, 
$g_{ij}(\boldsymbol{r},t)=\partial_i\vec n(\boldsymbol{r},t)\cdot\partial_j\vec n(\boldsymbol{r},t)/4$ is the quantum metric associated with the spin texture (i.e., of  the eigenstates $\ket{\psi_{\vec{r}}} \in \mathbb{C}^2$ of $\vec{n}(\vec{r},t)\cdot\vec{s}$), and summation over the repeated indices $i,j\in\{x,y\}$ is left implicit. 

Two comments about this result for the $d$-wave altermagnetic texture are in order: 
First, the itinerant electrons experience emergent orbital magnetic and electric fields, $\mathcal{B}(\boldsymbol{r},t)$ and $\mathcal{E}_i(\boldsymbol{r},t)$, as they move through the texture. These fields have opposite signs 
for the two sublattices due to the prefactor $s_{z}$ (which is equivalent to $-\tau_{z}$ in $P_-$) and require non-coplanar textures in space ($\mathcal{B}$) or space-time ($\mathcal{E}_i$).
However, we note that both fields are \textit{not} genuinely unique to altermagnets since they arise even in the antiferromagnet \cite{PhysRevB.86.245118} limit, $t_{z,\vec{p}}\rightarrow 0$, and as such not our main subject of interest here.
\begin{figure}[!t]
    \centering
    \includegraphics[width=1\linewidth]{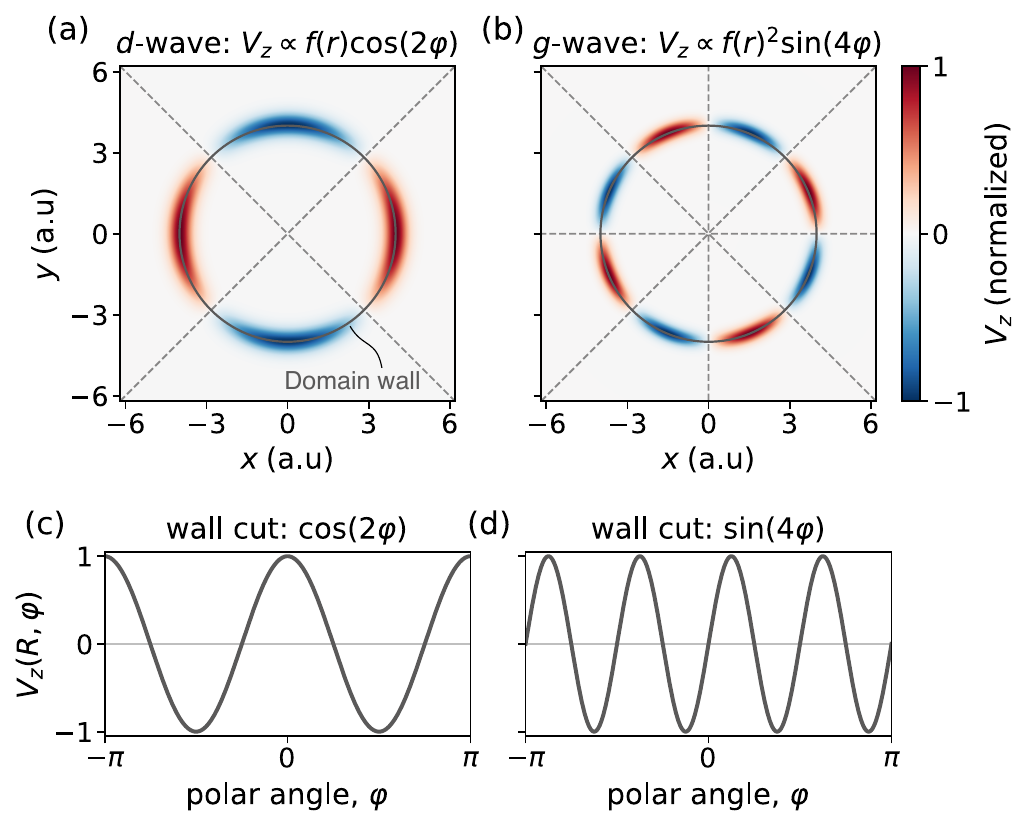}
    \caption{
\textbf{Multipolar emergent Zeeman fields from altermagnetic domain walls.}
(a) Emergent Zeeman field, $V_z(\boldsymbol{r})$ (normalized by its maximum), for a circular domain wall in a $d$-wave altermagnet, showing a quadrupolar pattern localized at the wall.
(b) Corresponding result for a $g$-wave altermagnet, showing an octupolar pattern.
(c,d) Angular cuts, $V_z(R,\varphi)$, at the wall radius, showing the $\cos(2\varphi)$ and $\sin(4\varphi)$ dependence.
The multipolar form of $V_z$ provides a novel probe for altermagnetic order.
    }
    \label{fig:2}
\end{figure}

Second, the itinerant electrons also experience an emergent Zeeman field, $V_{z}$. This Zeeman field is unique to the altermagnet, since it vanishes in the limit when $\varrho_{z}\rightarrow0$.
In particular, it is of geometric origin as it involves the quantum metric. 
To further explain the effects of this Zeeman field, we move to the principal axis frame of the quantum metric 
through a rotation, 
$
g = \mathcal{R}_{\psi}
\text{diag}(\lambda_{1},\lambda_{2})
\mathcal{R}_{\psi}^T
$ where $\mathcal{R}_{\psi}$ is a $2d$ rotation by an angle, $\psi$, and $\lambda_{1,2}\geq0$ are the eigenvalues of the quantum metric tensor.
Then, $V_{z}=(\hbar^2/4)(\lambda_1-\lambda_2)\cos 2\psi$. Hence, $V_{z}$ measures how the principal axes of the texture are oriented relative to $d$-wave altermagnetic order parameter. Such an effect does not occur for $V_0=(\hbar^2/4)(\lambda_1+\lambda_2)$, since it does not depend on $\psi$.

Another important feature of the emergent Zeeman field is that it produces a quadrupolar moment that inherits the 
the structure of the altermagnetic order through $\eta^{ij}$. To explain this feature, we consider a static radial domain wall with
$\theta(\boldsymbol{r})=\pi/2$
and 
$\phi(\boldsymbol{r})=(\pi/2)\tanh((r-R)/w)$ where $R$ is the domain radius, $w \ll R$ the domain wall width, and $r=|\boldsymbol{r}|$, see \figref{fig:1}. In this case, $\lambda_{1}=\phi'(r)^2$, $\lambda_{2}=0$, and $\psi=\varphi$ is the polar angle.
Hence, $V_{z}=f(r) \cos2\varphi$ with $f(r)=(\hbar^2/4)\phi'(r)^2$ realizes a ``four-lobe'' quadrupolar pattern near the domain wall boundary, which we illustrate in \figref{fig:2}(a,c). This quadrupole moment leads to spin accumulation near the domain wall.

An interesting question is whether such multipolar moments allow us to distinguish between different altermagnetic order parameters. To answer this question, we consider the previously introduced example of a $g$-wave altermagnet.
In this case, the effective Hamiltonian (again to $0^{\text{th}}$-order in $1/J$)
takes on the form,
\begin{equation}
\hat{\mathcal{H}}^{(0)}_{\text{rot}}
=
\varepsilon_{0,\hat{\boldsymbol{\Pi}}}\sigma_0 
+
({t}_{z,\hat{\boldsymbol{\Pi}}}+\delta{t}_{z,\hat{\boldsymbol{\Pi}}}) \sigma_z
+
\varrho_{0} V_0 \tau_0
+
\varrho_{z} V_z \sigma_z
\end{equation}
Here, $\delta{t}_{z,\hat{\boldsymbol{p}}}=6\hbar^2 \bar{\varrho}_z
T^{ijkl}
\mathcal W[\hat{p}_i\hat{p}_j g_{pl}(\boldsymbol{r},t)]$ is a correction term that is quadratic in the momenta, with
$
T^{ijkl}
=
\frac{1}{4!}\frac{1}{\bar{\varrho}_z}
\left.
\frac{\partial^4 t_{z,\boldsymbol{p}}}
{\partial k_i \partial k_j \partial k_k \partial k_l}
\right|_{\boldsymbol{p}=0}
$
denoting a totally symmetric rank-$4$ tensor
and $\mathcal W$ is the previously introduced Weyl-ordering.  
The emergent Zeeman field is given by
\begin{equation}
V_{z}
=
\hbar^4
T^{ijkl}
g_{ij}(\boldsymbol{r},t) g_{kl}(\boldsymbol{r},t).
\label{VsForgwave}\end{equation}

To contrast this emergent Zeeman field with the $d$-wave example, let us again consider a radial domain wall. In the $g$-wave case, we find that $V_z=(\hbar^4/64)\phi'(r)^4\sin4\varphi$, which realizes an eight-lobe octupolar pattern localized near the domain-wall boundary, see \figref{fig:2}(b,d). The multipolar pattern from the emergent Zeeman field thus indeed allows to distinguish between different altermagnetic order parameters. In general, we expect the octopolar pattern in the $g$-wave case to be weaker since $V_{z}\propto\hbar^4$, not $V_{z}\propto\hbar^2$ as in the $d$-wave case. However, it should still be relevant at sharper domain walls where $\phi'(r)^4$ is sizable.

\section{Geometry and electron dynamics}
So far, we have focused on the leading corrections in $1/J$ to describe the altermagnetic texture.  
At this zeroth order $1/J$, the quantum metric, $g_{ij}(\boldsymbol{r},t)$, enters only through the scalar potentials, $V_0$ and $V_z$.  
A natural question is if the metric can also modify the electron dynamics through the kinetic energy.

To address this question, we consider a time-independent $d$-wave altermagnetic texture.  
Let us begin by assuming that the inter-sublattice hybridization is the smallest energy scale and set $t_x=0$, which will allow for a systematic presentation of the findings and discussion of the different contributions.  
The first order correction to the effective Hamiltonian in the rotating frame is then $\hat{\mathcal{H}}^{(1)}_{\text{rot}}=-\Lambda^{\dagger}\Lambda/2J$,
where $\Lambda = P_- \hat{\mathcal H}_{\mathrm{rot}} P_+$ describes virtual transitions between the low-energy $S=-1$ and the high-energy $S=+1$ subspace. The full effective Hamiltonian up to first order in $1/J$ is given by
$\hat{\mathcal{H}}^{(\text{eff})}_{\text{rot}}=\hat{\mathcal{H}}^{(0)}_{\text{rot}}+\hat{\mathcal{H}}^{(1)}_{\text{rot}}$.

\begin{figure}[tb]
    \centering
    \includegraphics[width=1\linewidth]{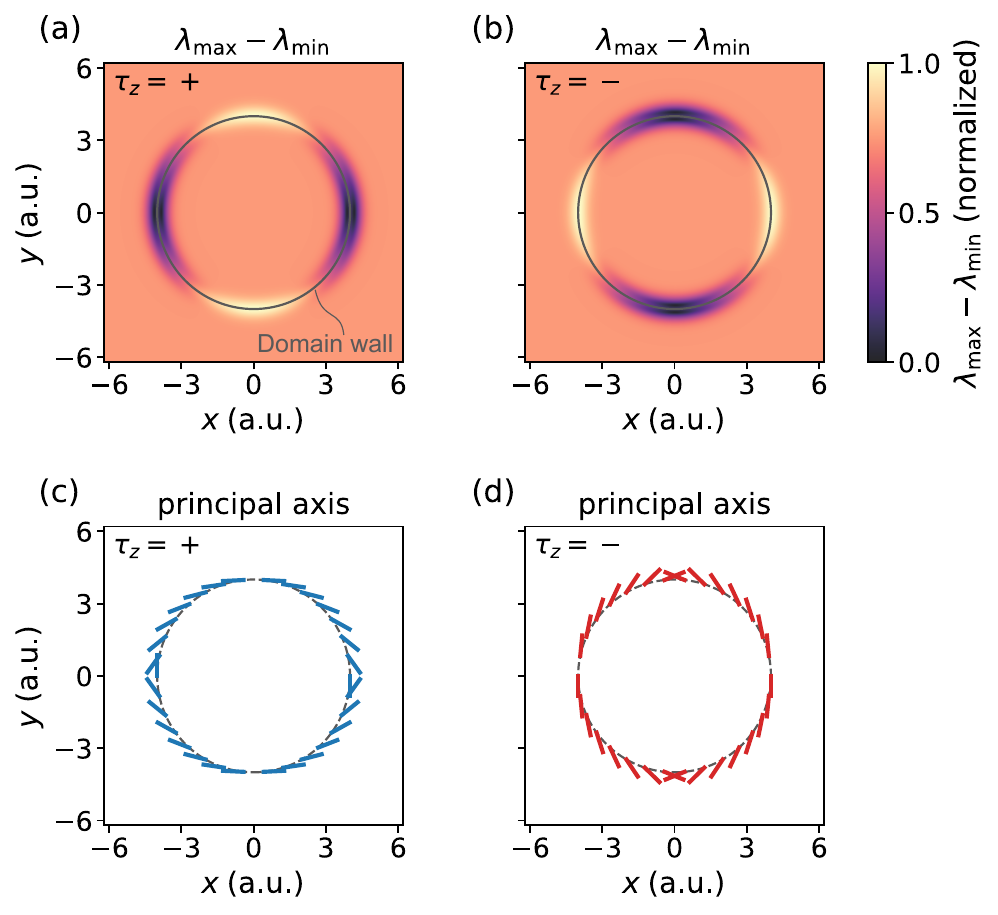}
\caption{\textbf{Texture-induced effective metric near a circular domain wall.}
(a,b) Spatial map of $\lambda_{\max}-\lambda_{\min}$ for $\tau_z=\pm$, where $\lambda_{\max/\min}(\boldsymbol{r})$ are the eigenvalues of the metric tensor, $g_{\text{eff},\tau}(\boldsymbol{r})$, that enters the kinetic energy.
At fixed $\boldsymbol{r}$, the local dispersion forms an ellipse in momentum space and $\lambda_{\max}-\lambda_{\min}$ quantifies its elongation.
(c,d) Major axis of the ellipse (eigenvector of $\lambda_{\max}$) evaluated along the wall, which sets the orientation of the local dispersion.
}
    \label{fig:3}
\end{figure}

We now evaluate $\hat{\mathcal H}^{(1)}_{\text{rot}}$ semiclassically; the full quantum mechanical operator expression is given in the Supplemental Material.  
Here we use a Wigner-Weyl formulation \cite{Curtright2014-pc}, which maps operators to phase-space functions and replaces operator products by Moyal products.
Keeping terms up to order $\hbar^0$ in the expansion of Moyal products corresponds to the semiclassical approximation. In this case, the correction to the Hamiltonian function is
\begin{equation}
\mathcal H^{(1)}_{\text{rot}}
=
-\frac{2\hbar^{2}}{J}\,
p_i 
[
K^{ij}
\,
g_{j\ell}(\boldsymbol{r})
\,
K^{\ell m}
]
p_m.  \label{VaryingG}
\end{equation}
Here, $K^{ij}=\varrho_0\delta^{ij}+\varrho_z\sigma_z\eta^{ij}$ is the kinetic-energy tensor of the $d$-wave altermagnet.

This result has a geometric interpretation.
For explaining this geometric interpretation, we focus on 
coplanar textures, so that the longitudinal gauge potential $A^{\parallel}_{i}(\boldsymbol{r})$ gives rise to a pure gauge and can be omitted.
For a fixed sublattice $\tau=\sigma=\pm$, the semiclassical spectrum is then given by, 
\begin{equation}
\varepsilon_\tau(\boldsymbol{p},\boldsymbol{r})
=
p_{i}
[g^{ij}_{{\text{eff}},\tau}(\boldsymbol{r})]
p_{j}
+
V_\tau(\boldsymbol{r}),
\end{equation}
with effective metric
$
g^{ij}_{\text{eff},\tau}(\boldsymbol{r})
=
K^{ij}_\tau
-
\frac{2\hbar^2}{J}
[
K_\tau g(\boldsymbol{r}) K_\tau
]^{ij}
$,
kinetic tensor
$
K^{ij}_\tau
=
\varrho_0 \delta^{ij}
+
\tau \varrho_z \eta^{ij}
$,
and total scalar potential
$
V_\tau
=
\varrho_0 V_0
+
\tau \varrho_z V_z
$.
Hence, the quantum metric gives rise to a sublattice-dependent ``deformation'' of the kinetic energy.
Equivalently, electrons propagate as if moving in a curved space whose geometry
is determined \textit{jointly}  by texture gradients (through $g_{ij}$) and by the altermagnetic order (through $\eta^{ij}$). 

The same geometric picture follows from the equations of motion.  
Treat $r_i$ and $p_{j}$ as phase-space variables with Poisson brackets 
$\{r_i,p_j\}=\delta_{ij}$.
Hamilton's equations are
$\dot r_i=\partial\varepsilon_\tau/\partial p_{i}$
and
$\dot{p}_{i,\tau}=-\partial\varepsilon_\tau/\partial r_i$.
Eliminating the canonical momenta leads to
\begin{equation}
\ddot r^{i}
+
\Gamma^{i}_{jk,\tau}(\boldsymbol{r})
\dot{r}^{j}
\dot{r}^{k}
=
-
2
g^{ij}_{{\text{eff}},\tau}
\partial_j V_\tau,
\label{Eq10}
\end{equation}
Here, the Christoffel symbols are 
$
\Gamma^{i}_{jk,\tau}(\boldsymbol{r})
=
\frac{1}{2}
(g_{\mathrm{eff},\tau})^{i\ell}
[
\partial_j g^{\text{eff}}_{\tau,k\ell}
+
\partial_k g^{\text{eff}}_{\tau,j\ell}
-
\partial_\ell g^{\text{eff}}_{\tau,jk}
]
$
with $g^{\text{eff}}_{\tau,ij}=(g^{-1}_{\text{eff},\tau})_{ij}$.  
Eq.\,\eqref{Eq10} shows that a semiclassical electron wavepacket is subject to the following effects: 
First, a force from the total scalar potential, $V_\tau$. Second, a purely geometric contribution encoded in the Christoffel symbols of the effective metric. 
Third, for noncoplanar textures, an additional Lorentz-like force, $\propto \epsilon_{jk}\mathcal B_\tau \dot r^k$ with opposite signs for the two $\tau$ sectors, where $\mathcal{B}_{\tau}=\tau \hbar\, \vec{n}\cdot[\partial_x\vec{n}\times\partial_y\vec{n}]/2$. This term vanishes for the coplanar textures considered here, but becomes finite when the spin texture acquires a scalar spin chirality.
All of these contributions differ significantly from their ferromagnetic analogue \cite{qxnw-8q4y} and also lead to crucial differences between altermagnets and antiferromagnets, which we will next discuss in the context of a circular domain wall inhomogeneity.

\section{Altermagnetic lensing}
The transport of electrons through the altermagnetic texture will provide a bending of electron trajectories, similar to the ``lensing effect'' in optics. We will focus on static, coplanar textures ($\mathcal{B}_{\tau}=0$ and $\mathcal{E}_{i}=0$), and discuss two cases: 
(1) lensing due to the scalar potential, $V_{\tau}$, (2) lensing due to effective metric, $g^{\text{eff}}_{\tau,ij}$. Here, (1) corresponds to the situation when $J$ is sufficiently large and electron speed is ``slow'', while (2) corresponds to fast electrons and moderately large $J$ so that metric corrections to the kinetic energy become relevant.
To see this, suppose we effectively increase the speed of electrons by rescaling the time coordinate, $t\rightarrow t/\alpha$ for some $\alpha\geq1$. Then, $\dot{r}\rightarrow\alpha\dot{r}$ and $\ddot{r}\rightarrow\alpha^2\ddot{r}$. Hence, Eq.\,\eqref{Eq10} reads 
$
\ddot{r}^{i}
+
\Gamma^{i}_{jk,\tau}
\dot{r}^{j}
\dot{r}^{k}
=
-
[2g^{ij}_{{\text{eff}},\tau}
\partial_j 
V_\tau]/\alpha^2$ and the effects of the scalar potential gets suppressed upon increasing $\alpha$ or, equivalently, the speed of the electrons~\cite{qxnw-8q4y}.

First, let us consider (1), lensing due to the scalar potential, $V_{\tau}$, and focus on the previously introduced radial domain wall example. 
In this case, $V_{z}(r)=f(r)\cos2\varphi$ gives rise to a force with a tangential component 
$
-\frac{1}{r}
\partial_\varphi V_z
=
\frac{2f(r)}{r}
\sin2\varphi
$ and opposite sign for the two sublattices. This force field is localized near the domain wall where $f(r)\neq0$. As a result an electron ray would propagate along a straight line away from the domain wall. At the wall, the ray receives a ``kick'' from the force field that leads to a change in its slope. This change in slope will depend on the sublattice/spin and the orientation of the ray relative to the crystal axis of the altermagnet; in particular, the 
``kick'' vanishes along the crystal axis when $\varphi=0,\pi/2$ and hence $\sin2\varphi=0$.

\begin{figure}[!t]
    \centering
    \includegraphics[width=1\linewidth]{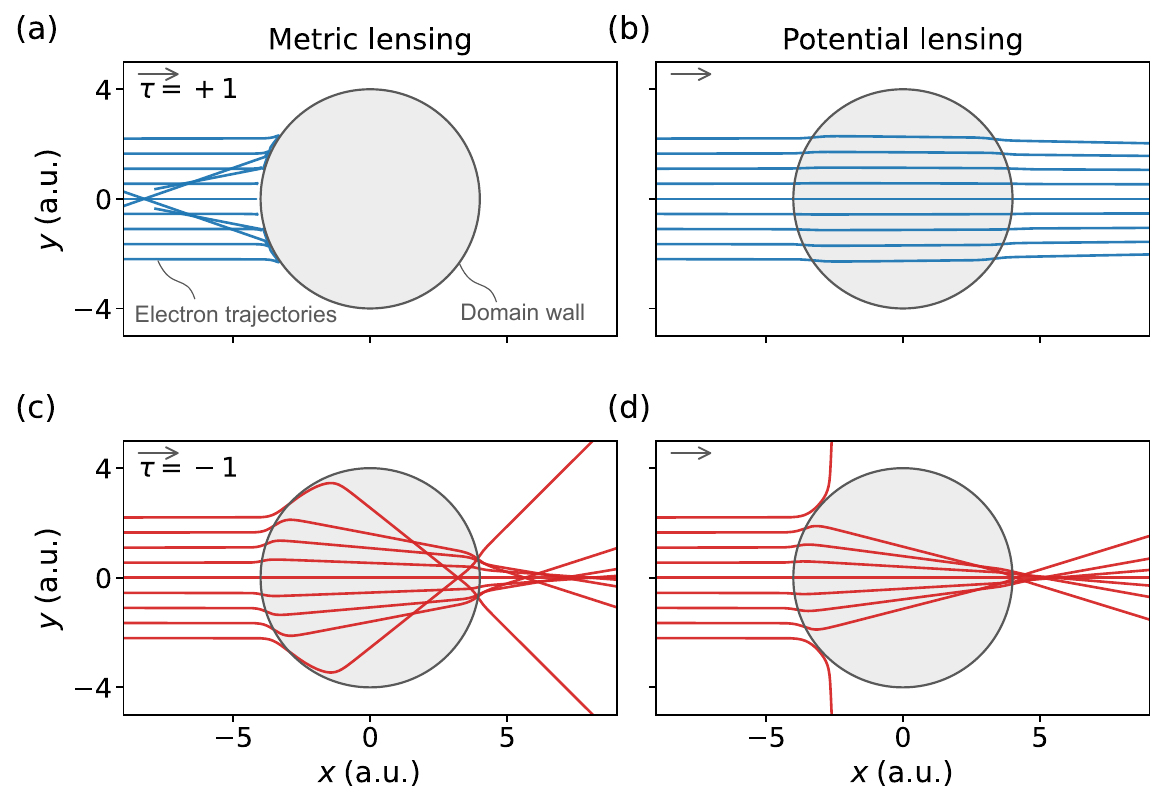}
    \caption{
\textbf{Altermagnetic domain wall as a sublattice-selective electron lens.}
Semiclassical trajectories incident from the left on a circular domain wall (gray disk), with $\tau=+1$ (blue) and $\tau=-1$ (red).
Left: metric lensing from the texture-induced effective metric $g^{\mathrm{eff}}_{\tau,ij}(\boldsymbol{r})$.
Right: potential lensing from the emergent scalar potential $V_\tau(\boldsymbol{r})$.
Potential lensing produces opposite deflections of the two $\tau$ sectors (focusing/defocusing). In contrast, metric lensing generates a transmission/reflection asymmetry.
    }
    \label{fig:4}
\end{figure}

To further illustrate this result, we can compute the momentum transfer by solving the equations of motion. Consider a ray injected with a group velocity
$
\boldsymbol v_{\text{in}}
\approx
\nabla_{\boldsymbol{p}}\varepsilon_{0,\boldsymbol{p}}
\big|_{\boldsymbol{p}=\boldsymbol{p}_{\text{in}}}
$
that hits the domain wall at a point $\boldsymbol{r}_0=R \hat{\boldsymbol{r}}(\varphi_0)$ 
where $\hat{\boldsymbol{r}}=(\cos\varphi,\sin\varphi)$ is the radial unit vector and $\varphi_0$ is he impact angle. The radial velocity component at the point of impact is $v_r
=
\boldsymbol v_{\text{in}}\cdot\hat{\boldsymbol{r}}(\varphi_0)$.
To obtain a compact analytical expression, let us first study our equations of motion in the limit of a thin domain wall, where $w$ is the shortest length scale. We find that the tangential momentum is deflected by an amount
\begin{equation}
\Delta p_{\varphi,\tau}
\approx \tau\frac{\hbar^2 \varrho_z \pi^2}{6 w R}
\frac{\sin 2\varphi_0}{v_r}. \label{AnalyticExpression}
\end{equation}
The deflection of the radial component is negligible, $\Delta p_{r,\pm}\approx 0$. This result shows, once more, that there is no deflection along the crystal axes, $\varphi_{0}=0,\pi/2$, and that it depends on the $\tau$ sector. What is more, we also see that the deflection gets negligible when $v_{r}$ becomes sizable. This corresponds to exactly the limit where effects of the quantum metric are of relevance.
We verified that this behavior is also qualitatively found when solving the semiclassical equations numerically for finite $w$, see \figref{fig:4}(b,d).

Let us continue with (2) and  consider the metric-induced lensing effects for the same radial domain wall example. In this case, the texture-induced quantum metric is 
$g_{ij}(\boldsymbol{r})
=
\phi'(r)^2
\hat{r}_{i}\hat{r}_{j}/4$
and the effective metric that the electrons feel is 
$
g^{ij}_{\text{eff},\tau}(\boldsymbol{r})
=
K_\tau^{ij}
-
\beta(r)
(K_{\tau}\hat{\boldsymbol{r}})^i
(K_{\tau}\hat{\boldsymbol{r}})^j
$
with
$
\beta(r)=(\hbar^2/2J)\phi'(r)^2.
$
The velocity, $\dot{\boldsymbol{r}}$, of an itinerant electron is determined by the conjugate momentum, $\boldsymbol{p}$, through the Hamilton 
equation 
$
\dot{\boldsymbol{r}}
=
2 g_{\text{eff},\tau}(\boldsymbol{r}) \boldsymbol{p}
$. Projecting this equation onto the radial direction, $\hat{\boldsymbol{r}}$, gives, 
\begin{equation}
\dot{r}
=
2
v^{(0)}_{g,r}(\varphi)
[
1-\beta(r)
\hat{\boldsymbol{r}}^{T}K_\tau \hat{\boldsymbol{r}}
]
\equiv
2
v^{(0)}_{g,r}
Z_\tau(r,\varphi) \label{EquationOfMotion}
\end{equation}
where 
$
v^{(0)}_{g,r}= \hat{\boldsymbol{r}}^T K_\tau \boldsymbol{p}
$
is the radial component of the group velocity of the uniform altermagnet
and 
$
Z_\tau(r,\varphi)
$
is a $\tau$-dependent transmission factor. It is defined through the second relation in \equref{EquationOfMotion} and can be explicitly written as 
$
Z_\tau(r,\varphi)
=
1-\beta(r)(\varrho_0+\tau \varrho_z\cos2\varphi)
$. This finding has interesting consequences by considering three different regimes:

If $Z_\tau>0$, then the velocity $\dot{r}$ has the same sign as 
$v^{(0)}_{g,r}$. Hence, the ray will move in the direction of the radial momentum projection.  
If $Z_\tau=0$, then $\dot{r}=0$ and the ray will be tangent to the wall.
If $Z_\tau<0$, then $\dot{r}$ has opposite sign as $v^{(0)}_{g,r}$.
Hence, the ray will be pushed away from the domain wall, even if the momentum direction points into the domain wall. Interestingly, due to the $\tau$-dependence of $Z_{\tau}$, a special situation can arise when, say, $Z_+>0$ but $Z_-<0$.
In this situation, the domain wall will act like a geometric spin filter, which we illustrate numerically in \figref{fig:4}(a,c). 
Only spins with one polarization are transmitted, while rays with the opposite polarization are scattered back. 
In terms of the microscopic system parameters the condition $Z_+>0$ but $Z_-<0$ is achieved when 
$
(\varrho_0-\varrho_z\cos2\varphi)^{-1}
< \beta
<
(\varrho_0+\varrho_z\cos2\varphi)^{-1}.
$
We emphasize that both spin-dependent scattering effects discussed in this section would be absent in the antiferromagnetic limit ($\varrho_z \rightarrow 0$) where neither $Z_\tau$ nor $V_\tau$ depend on $\tau$ [also note that $\Delta p_{\varphi,\tau} \propto \varrho_z$ in \equref{AnalyticExpression}].

\section{Emergent spin-orbit coupling and odd-parity magnetism}
We finally allow for finite $t_x$ such that there is an additional, non-trivial second-order process: the transverse gauge-field components $A^\perp$ change the spin of the electron and, subsequently, the $t_x$ term swaps the sublattice, bringing us back to the low-energy subspace (with $S=-1$). This time, however, the component of the spin has changed, leading to an additional contribution $\Delta \hat{\mathcal{H}}^{(1)}_{\text{rot}}$ to $\hat{\mathcal{H}}^{(1)}_{\text{rot}}$, which involves $\sigma_{x,y}$. While the full expression can be found in the Supplemental Material, the correction reads in the semi-classical limit as
\begin{subequations}\begin{equation}
    \Delta \mathcal{H}_{\text{rot}} = \vec{\mathtt{g}}_{\vec{p}}(\vec{r}) \cdot \vec{\sigma}_{\phi(\vec{r})}, \label{EffectiveSOC}
\end{equation}
with ``emergent spin-orbit field''
\begin{equation}
    \vec{\mathtt{g}}_{\vec{p}} = \mathtt{g}_0 \, p_j \begin{pmatrix}
        \sin (\theta) \partial_j \phi \\ \partial_j\theta
    \end{pmatrix}, \quad \mathtt{g}_0 = - \frac{\hbar \varrho_0 t_x}{J}, \label{gvector}
\end{equation}\label{EmergentSOCAllTerms}\end{subequations}
and rotated Pauli matrices $\vec{\sigma}_{\phi} = \mathcal{R}_{\phi} (\sigma_x,\sigma_y)^T$ where $\mathcal{R}_{\phi}$ rotates two-component vectors by an angle $\phi$. Note that the third second-order process, based on changing the sublattice twice with $t_x$, does not lead to a non-trivial operator in the low-energy subspace and, hence, will not be discussed here.

Importantly, it holds $\vec{\mathtt{g}}_{\vec{p}} = -\vec{\mathtt{g}}_{-\vec{p}}$ such that the term in \equref{EffectiveSOC} preserves time-reversal symmetry but is odd under inversion (at fixed position $\vec{r}$). As such, it can locally be thought of as effective spin-orbit coupling. However, this term does not arise from relativistic corrections but, instead, from non-trivial magnetic textures. Consequently, one might also think of \equref{EmergentSOCAllTerms} as representing some form of spontaneous generation of spin-orbit coupling \cite{PhysRevLett.93.036403,PhysRevB.75.115103,PhysRevB.95.125122} or an admixed odd-parity (anti-alter)magnetic component \cite{RafaelReview,2023arXiv230901607B}. We emphasize, however, that \equref{gvector} is not unique to altermagnets as it persists in the limit $\varrho_z \rightarrow 0$, as opposed to the lensing effects of the previous section.
\begin{figure}[!t]
    \centering
    \includegraphics[width=1.0\linewidth]{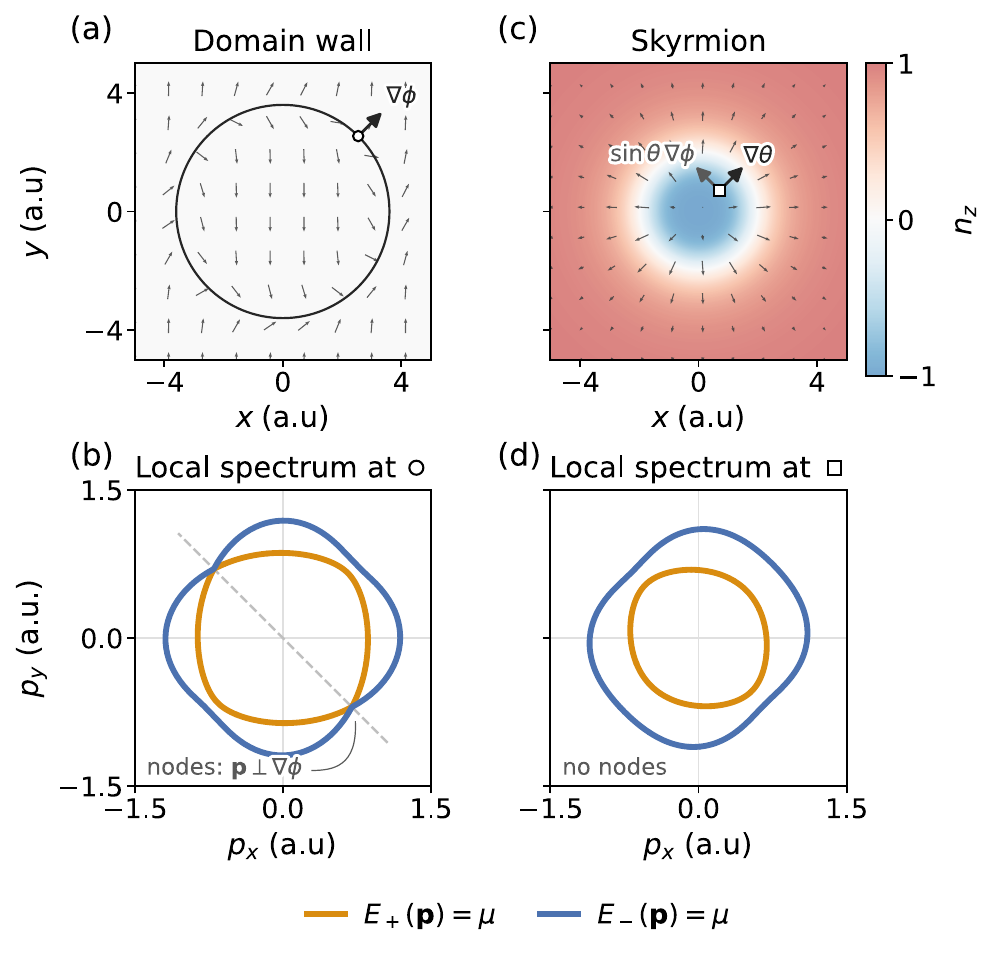}
    \caption{
\textbf{Emergent spin-orbit coupling and odd-parity magnetism from altermagnetic textures.}    
(a) Circular Néel domain wall. Faint arrows show the in-plane Néel vector field, $\boldsymbol{n}(\boldsymbol{r})$.
The marked point lies at azimuthal angle $\varphi=\pi/4$, where the emergent Zeeman field, $V_z \propto \cos 2\varphi$, vanishes.
(b) Local constant-energy contours, $E_\pm(\boldsymbol{p})=\mu$, evaluated at that point. Because $V_z=0$, the splitting arises purely from the metric-induced spin-orbit term, $\vec{\mathtt{g}}_{\boldsymbol{p}} \propto \boldsymbol{p}\cdot\nabla\phi$. 
The splitting vanishes for $\boldsymbol{p} \perp \nabla\phi$ (dashed line), producing a nodes.
(c) Skyrmion texture. The color scale shows $n_z$. Arrows indicate the in-plane components of $\boldsymbol{n}(\boldsymbol{r})$.
(d) Local constant-energy contours at the point in (c). Here, both $\nabla\theta$ and $\sin\theta\nabla\phi$ contribute to $\vec{\mathtt{g}}_{\boldsymbol{p}}$, which eliminates any nodes.
    }
    \label{fig:5}
\end{figure}

To discuss concrete examples, let us begin with co-planar textures, e.g., by setting $\theta = \pi/2$ but allowing for $\phi = \phi(\vec{r})$, which includes our domain wall example in \figref{fig:1} as a special case. From \equref{gvector}, we find $\vec{\mathtt{g}}_{\vec{p}}/\mathtt{g}_0 = (\vec{p}\cdot \vec{\nabla}\phi,0)^T$. As can be read off from the corresponding local spectrum
\begin{equation}
    E_\pm(\vec{p}) = \varepsilon_{0,\vec{p}} + \varrho_0 V_0 \pm \sqrt{(t_{z,\vec{p}}+ \varrho_z V_z)^2 + \vec{\mathtt{g}}_{\vec{p}}^2} \label{Dispersion}
\end{equation}
of our effective, semi-classical $2\times 2$ low-energy Hamiltonian $\mathcal{H}^{(\text{eff})}_{\text{rot}}$ (where we neglected the non-trivial metric in \equref{VaryingG} for simplicity), the spin-orbit splitting is $|\vec{\mathtt{g}}_{\vec{p}}|$ and thus finite except for $\vec{p}$ perpendicular to $\vec{\nabla}\phi$. In \figref{fig:5}(a,b), we illustrate this behavior for a domain wall, where $\vec{\nabla}\phi = \hat{\vec{r}} \phi'(r)$, by focusing on a single point at the boundary [indicated in \figref{fig:5}(b)]. Note that a planar spiral, i.e., $\phi(\vec{r}) = \vec{q}\cdot\vec{r}$, leads to the spatially constant spin-orbit field, $\vec{\mathtt{g}}_{\vec{p}} = (\vec{p}\cdot \vec{q},0)^T$, of the same form and, thus, the same type of splitting everywhere in space. We emphasize the similarity to the vanishing of the spin-splitting in one direction in the $p$-wave magnets discussed in \refcite{2023arXiv230901607B}.

Spatially varying altermagnets can also induce spin-orbit splittings without nodal directions in $\vec{\mathtt{g}}_{\vec{p}}$. This, however, necessitates that $\sin(\theta)\vec{\nabla} \phi$ and $\vec{\nabla} \theta$ are linearly independent, which in turn is equivalent to having a non-coplanar $\vec{n}(\vec{r})$. For instance, $\phi = \vec{q}_1 \cdot \vec{r}$ and $\theta = \vec{q}_2 \cdot \vec{r}$ yields a finite $|\vec{\mathtt{g}}_{\vec{p}}|$ for all momenta $\vec{p} \neq 0$ as long as $\vec{q}_1$ and $\vec{q}_2$ are linearly independent and for points in space with $\sin \vec{q}_2\cdot\vec{r} \neq 0$. Another important non-coplanar structure is a skyrmion, see \figref{fig:5}(c), which also gives rise to extended regions in space with non-nodal spin-orbit splitting, see \figref{fig:5}(d).

We finally point out that including an additional, subleading term in the intersublattice hopping, $t_{x,\vec{p}} = t_x + \varrho_x \vec{p}^2$, already generates a term of the same form as \equref{EmergentSOCAllTerms} at order $1/J^{0}$; now with $\mathtt{g}_0 = \hbar\varrho_3$. Which one is more relevant depends on how large $J$ is compared to $t_x \varrho_0/\varrho_x$.

\section{Conclusion}
Starting from the four-band model in \equref{Eq1}, with a slowly varying altermagnetic order parameter $\vec{n}(\vec{r},t)$, we have derived an effective two-band model describing the electronic spectrum in a rotating reference frame and studied the physical consequences of the different corrections that emerge relative to the homogeneous limit $\vec{n}(\vec{r},t) = \vec{n}_0$. First, to leading (zeroth) order $1/J$, we find that non-coplanar textures induce the emergent magnetic and electric fields given in \equref{EmergentEMFields}, which are opposite for the two sublattices and persist in the antiferromagnetic limit ($t_z \rightarrow 0$). We also uncover an emergent Zeeman field $V_z$, which, in contrast, is unique to the altermagnet. It involves the quantum metric $g_{ij}$ of the magnetic texture, and as such does not require non-coplanar order, as well as the momentum-space structure of the altermagnetic order parameter [via $\eta^{ij}$ and $T^{ijkl}$ in \equsref{VzFordwave}{VsForgwave} for a $d$-wave and $g$-wave altermagnet, respectively]. Since the associated local spin polarization can be probed experimentally, e.g., with scanning SQUIDs or NV magnetometry, this not only allows to distinguish altermagnets from antiferromagnets but also to access the predominant angular dependence of the altermagnetic order parameter [cf.~\figref{fig:2}(a) and (b)]. We have further shown that the emergent Zeeman field also leads to a spin-dependent lensing effect at inhomogeneities such as domain walls, see \figref{fig:4}(b,d).

At order $1/J$, additional corrections appear, which we divide into two categories. The first type preserves the spin/sublattice degree of freedom and can be thought of as an inhomogeneous, anisotropic contribution to the effective mass tensor or emulating curved space for electronic motion; as with the Zeeman field, it is unique to the altermagnet, involves the metric $g_{ij}$ and the form of the altermagnetic order parameter, and also introduces a spin-dependent lensing and, depending on parameters, filtering effect, see \figref{fig:4}(a,c). The second type of corrections to order $1/J$ mixes the spin species and can be interpreted as emergent spin-orbit coupling or the local admixture of an effectively odd-parity magnetic component. It can further gap out the nodes of the altermagnetic spin splitting of the bands.

Taken together, we have demonstrated that textures in altermagnets lead to particularly rich electronic properties. On the one hand, our findings show that taking into account such inhomogeneities, which are inevitable in experimental samples, is imperative for a proper understanding of the local and global electronic properties. On the other hand, altermagnetic textures also provide unprecedented opportunities for locally engineering electronic properties of interest, emulating dynamics in curved space and analogue gravity, creating spin-dependent electronic lenses and filters, and probing the altermagnetic order parameter symmetry experimentally.

\textit{Note added.} At the final stages of the preparation of this manuscript, \refcite{2026arXiv260214950M} appeared, which also discusses the impact of altermagnetic textures on electronic properties, albeit with a different focus.

\begin{acknowledgments}
M.S.S. thanks U.~Seifert for insightful discussions and acknowledges funding by the European Union (ERC-2021-STG, Project 101040651— SuperCorr). Views and opinions expressed are however those of the authors only and do not necessarily reflect those of the European Union or the European Research Council Executive Agency. Neither the European Union nor the granting authority can be held responsible for them. C.S. acknowledges support from the Louisiana Board of Regents.
\end{acknowledgments}

\onecolumngrid

\newpage

\clearpage
\setcounter{page}{1}
\renewcommand{\thepage}{\arabic{page}}

\setcounter{section}{0}
\renewcommand{\thesection}{S\arabic{section}}
\renewcommand{\thesubsection}{S\arabic{section}.\arabic{subsection}}
\renewcommand{\thesubsubsection}{S\arabic{section}.\arabic{subsection}.\arabic{subsubsection}}

\setcounter{figure}{0}
\setcounter{table}{0}
\renewcommand{\thefigure}{S\arabic{figure}}
\renewcommand{\thetable}{S\arabic{table}}

\setcounter{equation}{0}
\renewcommand{\theequation}{S\arabic{equation}}

\begin{center}
\large{\bf Supplemental Material \\}
\end{center}
\begin{center}Constantin Schrade$^{1}$ and Mathias S. Scheurer$^{2}$
\\
{\it $^{1}$Hearne Institute of Theoretical Physics, Department of Physics \& Astronomy, Louisiana State University, Baton Rouge LA 70803, USA}
\\
{\it $^{2}$Institute for Theoretical Physics III, University of Stuttgart, 70550 Stuttgart, Germany}
\end{center}

\section{Rotating Frame Hamiltonian}
In this section, we provide more details on the rotating frame Hamiltonian in which the quantization axis on each sublattice follows the local magnetization. For concreteness, we will focus on the case of a time-independent altermagnetic texture.
\\
\\
As a \textit{first step}, we redisplay (for convenience) the Hamiltonian for the altermagnetic spin texture, 
\begin{equation}
\hat{\mathcal{H}}
=
\hat\varepsilon_{0,\hat{\boldsymbol{p}}}\tau_0
+
\hat t_{x,\hat{\boldsymbol{p}}}\tau_x
+
\hat t_{z,\hat{\boldsymbol{p}}}\tau_z
+
J\tau_z\,\vec n(\boldsymbol{r})\cdot\vec{s} .
\end{equation}
as introduced in Eq.~(1) of the main text. We then parameterize the unit vector field $\boldsymbol{n}(\boldsymbol{r})$ as,
\begin{equation}
\boldsymbol{n}(\boldsymbol{r})
=
\left(
\sin\theta(\boldsymbol{r})\cos\phi(\boldsymbol{r}),
\sin\theta(\boldsymbol{r})\sin\phi(\boldsymbol{r}),
\cos\theta(\boldsymbol{r})
\right),
\end{equation}
and introduce a position-dependent spin rotation that aligns $\boldsymbol{n}(\boldsymbol{r})$ with the spin quantization axis,
\begin{equation}
U(\boldsymbol{r})
=
\exp\left[
-\frac{i}{2}\,\theta(\boldsymbol{r})
\left(
-\sin\phi(\boldsymbol{r})s_x+\cos\phi(\boldsymbol{r})s_y
\right)
\right],
\qquad
U^\dagger(\boldsymbol{r})\,
(\boldsymbol{s}\cdot\boldsymbol{n}(\boldsymbol{r}))\,
U(\boldsymbol{r})
=
s_z.
\end{equation}
\\
\\
As a \textit{second step}, we define the rotating-frame Hamiltonian as, 
\begin{equation}
\hat{\mathcal H}_{\mathrm{rot}}
\equiv
U^\dagger(\boldsymbol{r})\,
\hat{\mathcal H}\,
U(\boldsymbol{r}).
\end{equation}
which evaluates to, 
\begin{equation}
\hat{\mathcal H}_{\mathrm{rot}}
=
\hat\varepsilon_{0,\hat{\boldsymbol{\pi}}}\tau_0
+\hat{t}_{x,\hat{\boldsymbol{\pi}}}\tau_x
+\hat{t}_{z,\hat{\boldsymbol{\pi}}}\tau_z
+J\tau_z s_z,
\qquad
\hat{\boldsymbol{\pi}}
=
U^\dagger(\boldsymbol{r})\,\hat{\boldsymbol{p}}\,U(\boldsymbol{r})
=
\hat{\boldsymbol{p}}-\vec A(\boldsymbol{r}).
\end{equation}
We note that the spatial variation of the altermagnetic texture gives rise to emergent SU(2) gauge fields that couple to the electron momentum, 
\begin{equation}
A_{i}(\boldsymbol{r})
=
i\hbar\, U^\dagger(\boldsymbol{r}) \partial_i U(\boldsymbol{r}),
\qquad
\vec A(\boldsymbol{r})=(A_x(\boldsymbol{r}),A_y(\boldsymbol{r}),A_z(\boldsymbol{r})).
\end{equation}
For the parameterization introduced above, these SU(2) gauge fields take the explicit form,
\\
\\
As a \textit{third step}, it is useful to separate the SU(2) gauge fields into components that are parallel and perpendicular to the quantization axis,
\begin{equation}
A_i(\boldsymbol{r})=A_i^\parallel(\boldsymbol{r})+A_i^\perp(\boldsymbol{r}),
\qquad
A_i^\parallel \propto s_z,
\qquad
A_i^\perp \propto s_x,s_y,
\end{equation}
and thus commute and anticommute with $\tau_z s_z$, respectively.
These components are explicitly given by
\begin{equation}
\begin{split}
A_i^\parallel(\boldsymbol{r})
&=
-\hbar\,\partial_i\phi(\boldsymbol{r})\sin^2\left(\frac{\theta(\boldsymbol{r})}{2}\right)s_z
\\
A_i^\perp(\boldsymbol{r})
&=
\frac{\hbar}{2}
\left[
(\partial_i\theta(\boldsymbol{r}))\,\boldsymbol e_\phi(\boldsymbol{r})
-\sin\theta(\boldsymbol{r})\,(\partial_i\phi(\boldsymbol{r}))\,\boldsymbol e_r(\boldsymbol{r})
\right]\cdot(s_x,s_y),
\end{split}    
\end{equation}
with $\boldsymbol e_r(\boldsymbol{r})=
(\cos\phi(\boldsymbol{r}),\sin\phi(\boldsymbol{r}))^T$ and $\boldsymbol e_\phi(\boldsymbol{r})=
(-\sin\phi(\boldsymbol{r}),\cos\phi(\boldsymbol{r}))^T$.
\\
\\
Using this decomposition, we also introduce a momentum operator shifted by the longitudinal SU(2) gauge field,
\begin{equation}
\hat\Pi_i=\hat{p}_i-A_i^\parallel,
\end{equation}
so that the we can write 
\begin{equation}
\hat{\pi}_i=\hat\Pi_i-A_i^\perp.
\end{equation}
This formulation will be the starting point for our subsequent derivation of the low-energy effective Hamiltonian.

\section{Effective Hamiltonian}
In this section, we provide details on the derivations of the effective Hamiltonians shown in the main text.

\subsection{Schrieffer-Wolff Transformation}
We begin by redisplaying (for convenience) the rotating-frame Hamiltonian given in Eq.\,(2) of the main text, 
\begin{equation}
\hat{\mathcal H}_{\mathrm{rot}}
=
\hat\varepsilon_{0,\hat{\boldsymbol{\pi}}}\tau_0
+\hat t_{x,\hat{\boldsymbol{\pi}}}\tau_x
+\hat t_{z,\hat{\boldsymbol{\pi}}}\tau_z
+JS
\end{equation}
We then introduce the operator,
\begin{equation}
S \equiv \tau_z s_z,
\end{equation}
whose eigenvalues $s=\pm 1$ label the two subspaces separated by the large energy scale, $J$. In the rotating frame, $S$ is a good quantum number for distinguishing the low-energy ($s=-1$) and high-energy ($s=+1$) subspaces. The projection operators onto these subspaces are given by,
\begin{equation}
P_\pm = \frac{1}{2}(1 \pm S).
\end{equation}
It will now be convenient to decompose $\hat{\mathcal H}_{\mathrm{rot}}$ into diagonal and off-diagonal blocks with respect to the low- and high-energy subspaces,
\begin{equation}
H_{1,2}\equiv P_\mp \hat{\mathcal H}_{\mathrm{rot}} P_\mp,
\qquad
\Lambda \equiv P_- \hat{\mathcal H}_{\mathrm{rot}} P_+,
\end{equation}
With these definitions, we find that the low-energy effective Schrieffer-Wolff Hamiltonian is (to first order in $1/J$) given by,
\begin{equation}
H_{\mathrm{eff}}
=
H_1-\frac{1}{2J}\Lambda\Lambda^\dagger
\label{SWTrafo}\end{equation}
We will now evaluate this effective Hamiltonian for the $d$-wave and $g$-wave altermagnet.

\subsection{Effective Hamiltonian for the $d$-wave altermagnet (0$^{\text{th}}$ order in 1/J)}
For our first example of the $d$-wave altermagnet, we have 
$\hat\varepsilon_{0,\hat{\boldsymbol{p}}}=\varrho_0(\hat{p}_x^2+\hat{p}_y^2)$,
$\hat{t}_{x,\hat{\boldsymbol{p}}}=t_x$, 
and
$\hat t_{z,\hat{\boldsymbol{p}}}=\varrho_z(\hat{p}_x^2-\hat{p}_y^2)$. 
In this case, the kinetic term in our texture Hamiltonian can be written compactly as, 
\begin{equation}
\hat\varepsilon_{0,\hat{\boldsymbol{\pi}}}\tau_0+\hat t_{z,\hat{\boldsymbol{\pi}}}\tau_z
=
\hat\pi_i K^{ij}\hat\pi_j,
\quad \text{with} \quad
K^{ij}=\varrho_0\delta^{ij}+\varrho_z\tau_z\eta^{ij}
\quad \text{and} \quad
\eta=\mathrm{diag}(1,-1).
\end{equation}
\\
\\
We initially decompose the SU(2) gauge fields as $\hat\pi_i=\hat\Pi_i-A_i^\perp$ with $\hat\Pi_i=\hat k_i-A_i^\parallel$ and rewrite the first contribution to the kinetic term as, 
\begin{equation}
\begin{split}
\hat{\pi}_iK^{ij}\hat{\pi}_j
&=
(\hat{\Pi}_i-A_i^\perp)K^{ij}(\hat{\Pi}_j-A_j^\perp)
\\
&=
\hat{\Pi}_iK^{ij}\hat{\Pi}_j
-
(
\hat{\Pi}_iK^{ij}A_j^\perp
+A_i^\perp K^{ij}\hat{\Pi}_j
)
+A_i^\perp K^{ij}A_j^\perp
\\
&
=
\hat{\Pi}_iK^{ij}\hat{\Pi}_j
-
(
\hat{\Pi}_iK^{ij}A_j^\perp
+K^{ij}A_j^\perp\hat{\Pi}_i
)
+A_i^\perp K^{ij}A_j^\perp \\
&=
\hat{\Pi}_iK^{ij}\hat{\Pi}_j
-\{\hat{\Pi}_i,\;K^{ij}A_j^\perp\}
+A_i^\perp K^{ij}A_j^\perp,
\end{split}
\end{equation}
Using this result, the full rotating-frame Hamiltonian takes on the form, 
\begin{equation}
H_{\text{rot}}
=
\big[
\hat{\Pi}_iK^{ij}\hat{\Pi}_j
+A_i^\perp K^{ij}A_j^\perp
\big]
-\{\hat{\Pi}_i,\;K^{ij}A_j^\perp\}
+t_x\tau_x
+JS.
\end{equation}
\\
\\
With this rewriting, we are now in the position to evaluate the lowest-order contribution to our effective Hamiltonian, 
\begin{equation}
\begin{split}
H^{(0)}_{\text{rot}} &= H_1 \\
&
=
- J  +
P_-[\hat{\Pi}_iK^{ij}\hat{\Pi}_j
+A_i^\perp K^{ij}A_j^\perp  ]P_-
\\
&=
-J
+
P_-
[
\varrho_0 (\hat{\Pi}_x^2+\hat{\Pi}_y^2)
+
\varrho_z\tau_z(\hat{\Pi}_x^2-\hat{\Pi}_y^2)
+
\big(\varrho_0V_{1}+\varrho_z\tau_z V_z\big)
]
P_-,
\end{split}    
\end{equation}
where we have defined the quantities,
\begin{equation}
\begin{split}
V_{0}
&=
\frac{\hbar^2}{4}
[
(\partial_x\theta)^2+(\partial_y\theta)^2
+
\sin^2\theta\big((\partial_x\phi)^2+(\partial_y\phi)^2\big)
]
=
\hbar^2
\delta^{ij}\,
g_{ij}(\boldsymbol{r}),
\\
V_z
&=
\frac{\hbar^2}{4}
[
(\partial_x\theta)^2-(\partial_y\theta)^2
+
\sin^2\theta\big((\partial_x\phi)^2-(\partial_y\phi)^2\big)
]
=
\hbar^2
\eta^{ij}\,
g_{ij}(\boldsymbol{r}),
\end{split}    
\end{equation}
with $\eta=\mathrm{diag}(1,-1)$ and the quantum metric $g_{ij}(\boldsymbol{r})
=
\partial_i \vec{n}(\boldsymbol{r})\cdot
\partial_j \vec{n}(\boldsymbol{r})/4$. 
Introducing the Pauli matrices $\sigma_j$ that act in the low-energy subspace, with basis $(\tau=+,s=\downarrow)$ and $(\tau=-,s=\uparrow)$, we obtain the effective low-energy Hamiltonian given by (upon dropping an irrelevant constant)
\begin{equation}
\begin{split}
H^{(0)}_{\text{rot}} =
\varrho_0 \sigma_0 (\hat{\Pi}_x^2+\hat{\Pi}_y^2)
+
\varrho_z\sigma_z(\hat{\Pi}_x^2-\hat{\Pi}_y^2)
+
\big(\varrho_0V_{1}+\varrho_z\sigma_z V_z\big).
\label{ProjectedHamZero}\end{split}    
\end{equation}
This concludes our derivation for the effective Hamiltonian of the $d$-wave altermagnet system.

\subsection{Effective Hamiltonian for the $g$-wave altermagnet (0$^{\text{th}}$ order in 1/J)}
For our second example of the $g$-wave altermagnet, we again set 
$\hat\varepsilon_{0,\hat{\boldsymbol{p}}}=\varrho_0(\hat p_x^2+\hat p_y^2)$,
$\hat t_{x,\hat{\boldsymbol{p}}}=t_x$. Moreover, we conveniently write, 
\begin{equation}
\hat t_{z,\hat{\boldsymbol{\pi}}}
=
\bar{\varrho}_z
\mathcal W[\hat{\pi}_x \hat{\pi}_y(\hat{\pi}_x^2-\hat{\pi}_y^2)]
=
\bar{\varrho}_z
T^{ijkl}
\mathcal W[\hat\pi_i\hat\pi_j\hat\pi_k\hat\pi_l] .
\end{equation}
where we have defined the totally symmetric rank-4 tensor $T^{ijkl}$ as,
\begin{equation}
T^{ijkl}
=
\frac{1}{4!}
\frac{1}{\bar{\varrho}_z}
\left.
\frac{\partial^4 t_{z,\boldsymbol{p}}}
{\partial{p}^i\partial{p}^j\partial{p}^k\partial{p}^l}
\right|_{\boldsymbol{p}=0} .
\end{equation}
In particular, we remark that its non-zero components are given by,
\begin{equation}
T^{xxxy}=T^{xxyx}=T^{xyxx}=T^{yxxx}=\frac{1}{4},
\qquad
T^{xyyy}=T^{yxyy}=T^{yyxy}=T^{yyyx}=-\frac{1}{4},
\end{equation}
As a next step, we insert our decomposition of the shifted momenta in terms of the longitudinal and transversal SU(2) gauge fields, 
$\hat\pi_i=\hat\Pi_i-\hat A_i^\perp$
and
$
\hat\Pi_i=\hat k_i-\hat A_i^\parallel
$. Projected onto the $S=-1$ subspace, we find that 
\begin{equation}
\begin{split}
P_-
\hat t_{z,\hat{\boldsymbol{\pi}}}
P_-
&=
\bar{\varrho}_z
P_-
T^{ijkl}
\left[
\mathcal W[\hat\Pi_i\hat\Pi_j\hat\Pi_k\hat\Pi_l]
+
6\,\mathcal W[\hat\Pi_i \hat\Pi_j \hat A_k^\perp \hat A_l^\perp]
+
\mathcal W[
\hat A_i^\perp
\hat A_j^\perp
\hat A_k^\perp
\hat A_l^\perp
]
\right]
P_-
\end{split}
\end{equation}
where we have used that only products involving an even number of transversal gauge field factors are non-zero 
upon projecting with $P_{-}$. 
\\
\\
To further simplify the above expression, we note that 
\begin{equation}
\begin{split}    
\hat A_k^\perp \hat A_l^\perp
&=
\frac{1}{2}
\{
\hat A_k^\perp,\hat A_l^\perp
\}
+
\frac{1}{2}
[\hat A_k^\perp,\hat A_l^\perp]
\\
&=
\hbar^2
(
g_{ij}
+ i\,\chi_{ij}\,s_z
)
\label{AAProduct}\end{split}
\end{equation}
where 
$\chi_{ij}
=
\vec n \cdot
(\partial_i \vec n \times \partial_j \vec n)/4$
is the scalar spin chirality. This result can be seen
from the explicit form of the transverse gauge potential, 
$
\hat A_i^\perp=\frac{\hbar}{2}
(
(\partial_i\theta)\mathbf e_\phi
-
\sin\theta\,(\partial_i\phi)\mathbf e_r
)
\cdot(s_x,s_y)
$, and the quantum metric,
$
g_{jk}(\mathbf r)
=
[
(\partial_j\theta)(\partial_k\theta)
+
\sin^2\theta\,(\partial_j\phi)(\partial_k\phi)
]/4
= (\partial_i \vec n) \cdot (\partial_j \vec n)/4$.
\\
\\
We can now write the second term in our expansion of 
$P_-
\hat t_{z,\hat{\boldsymbol{\pi}}}
P_-$ as, 
\begin{equation}
\begin{split}
T^{ijkl}
\mathcal{W}[\hat\Pi_i \hat\Pi_j \hat A_k^\perp \hat A_l^\perp]
&=
\frac{1}{2}
T^{ijkl}
\mathcal{W}[\hat\Pi_i \hat\Pi_j[\hat A_k^\perp,\hat A_l^\perp]]
+
\frac{1}{2}
T^{ijkl}
\mathcal{W}[\hat\Pi_i \hat\Pi_j\{\hat A_k^\perp,\hat A_l^\perp\}]
\\
&=
\frac{1}{2}
T^{ijkl}
\mathcal{W}[\hat\Pi_i \hat\Pi_j\{\hat A_k^\perp,\hat A_l^\perp\}]
\\
&=
\hbar^2
T^{ijkl}
\mathcal{W}[\hat\Pi_i \hat\Pi_j\, g_{kl}(\mathbf r)] .
\end{split}
\end{equation}
Here, we have noted that 
$T^{ijkl}$ 
and 
$\{
\hat A_k^\perp,\hat A_l^\perp
\}$ 
are both even under 
$k\leftrightarrow l$, 
while 
$[\hat A_k^\perp,\hat A_l^\perp]$ 
is odd under $k\leftrightarrow l$ and thus vanishes upon carrying out the summation. 
\\
\\
In a similar way, we find for the last term in the expansion of $P_-
\hat t_{z,\hat{\boldsymbol{\pi}}}
P_-$ that, 
\begin{equation}
\begin{split}
T^{ijkl}
\hat A_i^\perp
\hat A_j^\perp
\hat A_k^\perp
\hat A_l^\perp
&=
\hbar^4 T^{ijkl}
[
(
g_{ij} g_{kl}
-
\chi_{ij} \chi_{kl}
)
+
i
(
g_{ij}\chi_{kl}
+
\chi_{ij} g_{kl}
)s_z
]
=
\hbar^4 T^{ijkl} g_{ij} g_{kl},
\end{split}    
\end{equation}
where the last equality follows because $T^{ijkl}$ is totally symmetric and  $\chi_{ij}$ is antisymmetric under $i \leftrightarrow j$. 
\\
\\
We can now combine these results to arrive at, 
\begin{equation}
\begin{split}
\hat{\mathcal{H}}^{(0)}_{\text{rot}}
&=
P_-
[
\hat\varepsilon_{0,\hat{\boldsymbol{\Pi}}}\tau_0 
+
(\hat{t}_{z,\hat{\boldsymbol{\Pi}}}+\delta\hat{t}_{z,\hat{\boldsymbol{\Pi}}}) \tau_z
+
\varrho_0 V_0 \tau_0
+
\bar{\varrho}_z V_z \tau_z
]
P_{-},
\end{split}    
\end{equation}
with
\begin{equation}
\begin{split}
 \delta\hat{t}_{z,\hat{\boldsymbol{p}}}&=6\hbar^2 \bar{\varrho}_z 
T^{ijkl}
\mathcal W[\hat{p}_i\hat{p}_j g_{kl}(\boldsymbol{r},t)]   
\\
V_{z}
&=
\hbar^4
T^{ijkl}g_{ij}(\boldsymbol{r},t) g_{kl}(\boldsymbol{r},t), 
\end{split}    
\end{equation}
This concludes our derivation of the effective Hamiltonian to $0^{\text{th}}$ order in $1/J$ for the $g$-wave altermagnet example.

\subsection{Effective Hamiltonian for the $d$-wave altermagnet (1$^{\text{st}}$ order in 1/J)}
As a next step, we return to our first first example of the $d$-wave altermagnet and compute the
correction to the effective Hamiltonian in 1$^{\text{st}}$ order in $1/J$. 
\\
\\
For this purpose, it is initially useful to formulate a few definitions and identities. The \textit{first identity} is 
\begin{equation}
A_i^\perp A_j^\perp
=
\hbar^2
(g_{ij} +
i
\chi_{ij}
s_z
)
\quad
\text{with}
\quad 
\{
A_i^\perp,A_j^\perp
\}
=
2\hbar^2 
g_{ij}
\quad
\text{and}
\quad 
[A_i^\perp,A_j^\perp]
=
2i\hbar^2\chi_{ij}s_z
\end{equation}
with $\chi_{ij}
=
\vec{n}\cdot(\partial_i\vec{n}\times\partial_j\vec{n})/4$. 
The \textit{second identity} is 
\begin{equation}
g_{ij}=\sum_{a=x,y }e_{ia} e_{ja},
\end{equation}
where,
\begin{equation}
\begin{split}
A_i^\perp=\hbar\sum_{a=x,y}e_{ia}s_a,
\quad
\text{with}
\quad
e_{ix}
&=
-
\frac{1}{2}\left[(\partial_i\theta)\sin\phi
+\sin\theta(\partial_i\phi)\cos\phi\right],
\\
e_{iy}
&=
\frac{1}{2}\left[(\partial_i\theta)(\cos\phi)
-\sin\theta(\partial_i\phi)\sin\phi\right].
\label{eiDefinitions}\end{split}
\end{equation}
The \textit{third identity} is, 
\begin{equation}
\{\hat\Pi_i,A_j^\perp\}
=
\{\hat p_i,A_j^\perp\}
=
\hbar\sum_{a=x,y}\{\hat p_i,e_{ja}\}s_a
\end{equation}
where we used that $\{s_{z},A_j^\perp\}=0$ in the first equality. 
\\
\\
With the help of these identities, we can now compute the 1$^{\text{st}}$ order correction in $1/J$ to the effective Hamiltonian. For this purpose, we write the off-diagonal blocks as, 
\begin{equation}
\Lambda
=
P_-H_{\mathrm{rot}}P_+
=
P_-\big(
t_x\tau_x
-\{\hat{\Pi}_i,\;K^{ij}A_j^\perp\}
\big)P_+
\end{equation}
and define,
\begin{equation}
\Lambda=\Lambda_\Delta+\Lambda_\pi,
\qquad
\Lambda_\Delta \equiv P_-(t_x\tau_x)P_+,
\qquad
\Lambda_\pi \equiv -P_-\{\hat{\Pi}_i,\;K^{ij}A_j^\perp\}P_+
\end{equation}
We then decompose the following product, 
\begin{equation}
\Lambda\Lambda^\dagger
=
\Lambda_\Delta\Lambda_\Delta^\dagger
+\Lambda_\pi\Lambda_\pi^\dagger
+\Lambda_\Delta\Lambda_\pi^\dagger
+\Lambda_\pi\Lambda_\Delta^\dagger.
\end{equation}
as it appears in our effective Hamiltonian and evaluate this expression term by term:
\\
\\
First, we consider the contribution originating from the sublattice hybridization. This term takes the form,
\begin{equation}
\Lambda_\Delta\Lambda_\Delta^\dagger
=
t_x^2 P_-,
\end{equation}
which gives a constant energy shift within the low-energy ($S=-1$) subspace.
\\
\\
Second, we consider the product $\Lambda_\pi\Lambda_\pi^\dagger$. Here, we first write,
\begin{equation}
\begin{split}
\Lambda_\pi
&=
-P_-\{\hat\Pi_i,K^{ij}A_j^\perp\}P_+
=
-\hbar K^{ij}\sum_{a=x,y}\{\hat p_i,e_{ja}\} (P_-s_aP_+)
\\
\Lambda_\pi^\dagger
&=
-\hbar
K^{k\ell}
\sum_{a'=x,y}
\{\hat p_k,e_{\ell a'}\}\,(P_+s_{a'}P_-). 
\end{split}
\end{equation}
We then find that the product of these operators can be written as, 
\begin{equation}
\begin{split}
\Lambda_\pi\Lambda_\pi^\dagger
&=
\hbar^2
K^{ij}K^{k\ell}
\sum_{a,a'=x,y}
\{\hat p_i,e_{j a}\}
\{\hat p_k,e_{\ell a'}\}
\left(P_- s_a P_+ s_{a'} P_-\right)
\\
&=
\hbar^2
K^{ij}K^{k\ell}
\sum_{a,a'=x,y}
\{\hat p_i,e_{j a}\}
\{\hat p_k,e_{\ell a'}\}
P_- (s_as_{a'}) P_- 
\\
&=
\hbar^2 K^{ij}K^{k\ell}
\left[
\sum_{a=x,y}
\{\hat p_i,e_{j a}\}\{\hat p_k,e_{\ell a}\}\,P_- 
+
i\sum_{a,a'=x,y}
\varepsilon_{aa'}
\{\hat p_i,e_{j a}\}\{\hat p_k,e_{\ell a'}\}
\,(P_- s_z P_-)
\right]
\end{split}
\end{equation}
where we have used that 
$
P_- s_a P_+ s_{a'} P_-
=
P_- s_as_{a'} P_- 
$
and, for $a,a'=x,y$, we have
$
s_as_{a'}
=
\delta_{aa'}
+
i\varepsilon_{aa'}s_z
$
with the totally antisymmetric tensor $\varepsilon_{aa'}$ that satisfies $\varepsilon_{xy}=1$.
\\
\\
In the semiclassical limit, we find for the first term in the above expression,
\begin{equation}
\Lambda_\pi\Lambda_\pi^\dagger
\rightarrow
4\hbar^2
(p_iK^{ij})\,g_{j\ell}(K^{\ell m}p_m).
\end{equation}

Finally, we evaluate the cross terms involving both $\Lambda_\Delta$ and $\Lambda_\pi$. 
\begin{equation}
\Lambda_\Delta
\Lambda_\pi^\dagger
+
\Lambda_\pi
\Lambda_\Delta^\dagger
=
-t_x
P_-
\left\{
\tau_x,
\{\hat\Pi_i,K^{ij}A_j^\perp\}
\right\}
P_- .
\end{equation}
where we have used that for any pair of operators, $A$ and $B$, that anti-commute with $S$, we have
$
(P_-
A
P_+)
(P_+
B
P_-)
+
(P_-
B
P_+)(
P_+
A
P_-)
=
P_-\{A,B\}P_- 
$ with $O_1=t_x\tau_x$ and $O_2=-\{\hat\Pi_i,\,K^{ij}A_j^\perp\}$.
\\
\\
To further simplify this, we note the general anti-commutator identity
$
\{A,\{B,C\}\}
=
\{B,\{A,C\}\}
$
if
$[A,B]=0$. Using this identity combined with $[\tau_x,\hat{\Pi}_i]=0$ and $[\tau_x,A_j^\perp]=0$ (since $\hat{\Pi}_i$ and $A_j^\perp$ are comprised of $s_{x,y,z}$ Pauli matrices), we find that,
\begin{equation}
\begin{split}
\Lambda_\Delta
\Lambda_\pi^\dagger
+
\Lambda_\pi
\Lambda_\Delta^\dagger
&=
-t_x
P_-
\left\{
\tau_x,
\{\hat\Pi_i,K^{ij}A_j^\perp\}
\right\}
P_- 
\\
&=
-t_x
P_-
\left\{
\hat\Pi_i,
\{\tau_x,K^{ij}A_j^\perp\}
\right\}
P_- 
\\
&=
-t_x
P_-
\left\{
\hat\Pi_i,
\{\tau_x,K^{ij}\}A_j^\perp
\right\}
P_- 
\\
&=
-t_x
P_-
\left\{
\hat{\Pi}_i,
\{\tau_x,K^{ij}\}A_j^\perp
\right\}
\\
&=
-t_x
P_-
\left\{
\hat{\Pi}_i,
\{\tau_x,K^{ij}\}A_j^\perp
\right\}
P_- 
\\
&=
-2\varrho_0 t_x 
\sum_{i=x,y}
P_-
\{\hat\Pi_i,\tau_x A_i^\perp\}
P_- 
\\
&=
-2\varrho_0 t_x 
\sum_{i=x,y}
P_-
\tau_x\{\hat\Pi_i, A_i^\perp\}
P_- 
\\
&=
-2\varrho_0 t_x 
\sum_{i=x,y}
P_-
\tau_x\{\hat{p}, A_i^\perp\}
P_- 
\\
&=
-2\hbar \varrho_0 t_x
\sum_{i=x,y}
\sum_{a=x,y}
\{\hat \Pi_i,e_{ia}(\boldsymbol{r})\}
P_-(\tau_xs_a)\,P_-
\label{MixedContribution}\end{split}\end{equation}
Here, we also used $\{A_i^\parallel,A_i^\perp\}=0$ and 
$
\{\tau_x,K^{ij}\}
=
\{\tau_x,\varrho_0\delta^{ij}+\varrho_z\tau_z\eta^{ij}\}
=
2\varrho_0\delta^{ij}\tau_x.
$
In summary, with this we see that the full quantum mechanical correction is given by,
\begin{equation}
\begin{split}
\hat{\mathcal H}^{(1)}_{\text{rot}}
&=
-\frac{1}{2J}
\Big[
t_x^{2}P_-
+
\hbar^2\,K^{ij}K^{k\ell}
\Big(
\sum_{a=x,y}
\{\hat p_i,e_{ja}(\boldsymbol{r})\}
\{\hat p_k,e_{\ell a}(\boldsymbol{r})\}
P_-
+
i\sum_{a,a'=x,y}
\varepsilon_{aa'}
\{\hat p_i,e_{ja}(\boldsymbol{r}\}\{\hat p_k,e_{\ell a'}(\boldsymbol{r})\}\,
(P_-s_zP_-)
\Big)
\\
&\quad-2\hbar \varrho_0 t_x 
\sum_{i=x,y}
\sum_{a=x,y}
\{\hat{p}_i,e_{ia}(\boldsymbol{r})\}
P_-(\tau_xs_a)P_-
\Big]
\end{split}
\end{equation}
We will now discuss the term $\propto \varrho_0t_{x}$ in more detail.
\\
\\
Defining, as before, Pauli matrices $\sigma_j$ acting in the low-energy subspace ($P_-$ projects on) with basis choice $(\tau=+,s=\downarrow)$ and $(\tau=-,s=\uparrow)$, the associated contribution to the Hamiltonian reads as
\begin{equation}
    \Delta\hat{\mathcal{H}}_{\text{rot}} = \frac{\hbar \varrho_0 t_x}{J} \sum_{i=x,y} \left( \{\hat{p}_i, e_{ix}(\vec{r})\}\sigma_x - \{\hat{p}_i, e_{iy}(\vec{r})\}\sigma_y \right), \label{SOCTermInduced}
\end{equation}
where we explicitly indicated that $e_{ia}(\vec{r})$, as defined in \equref{eiDefinitions}, depend in general on position. Again in the semi-classical limit,  
we can effectively replace $\{\hat{p}_i, e_{ia}(\vec{r})\} \rightarrow 2p_i e_{ia}$ leading to
\begin{equation}
    \Delta \mathcal{H}_{\text{rot}} = \frac{2\hbar \varrho_0 t_x}{J} \sum_{i=x,y} p_i \left( e_{ix}(\vec{r})  \sigma_x - e_{iy}(\vec{r})  \sigma_y \right) = \vec{\mathtt{g}}_{\vec{p}}(\vec{r}) \cdot \vec{\sigma}_{\phi(\vec{r})},
\end{equation}
where we introduced the spatially varying spin-orbit field and rotated Pauli matrices given by (suppressing position dependence for notational simplicity)
\begin{equation}
    \vec{\mathtt{g}}_{\vec{p}} = - \frac{\hbar \varrho_0 t_x}{J} \begin{pmatrix} \sin (\theta) \, \vec{p} \cdot \vec{\nabla}\phi \\ \vec{p} \cdot \vec{\nabla}\theta \end{pmatrix}, \quad \text{and} \quad   \vec{\sigma}_{\phi} = \mathcal{R}_{\phi} \begin{pmatrix} \sigma_x \\ \sigma_y \end{pmatrix}, \quad \mathcal{R}_{\phi} =  \begin{pmatrix}
        \cos \phi & - \sin \phi \\ \sin \phi & \cos \phi
    \end{pmatrix},
\end{equation}
respectively. 

To derive the condition stated in the main text, we first note that 
\begin{equation}
    \vec{\mathtt{g}}^2_{\vec{p}} = \frac{\hbar^2 \varrho_0^2 t^2_x}{J^2} \left[ \sin^2\theta (\vec{p}\cdot \vec{\nabla}\phi)^2 + (\vec{p} \cdot \vec{\nabla}\theta)^2 \right]
\end{equation}
is non-vanishing for all $\vec{p} \neq 0$ if and only if $\sin\theta \neq 0$ and the two vectors $\vec{\nabla}\phi$ and $\vec{\nabla}\theta$ are linearly independent. At the same time, the texture is (locally) non-collinear if and only if
\begin{align}
    \vec{n} \cdot \left( \partial_x \vec{n} \times \partial_y \vec{n} \right) &= \sin \theta \left[ (\partial_x\theta) (\partial_y \phi) - (\partial_y\theta) (\partial_x \phi) \right] \\ &= \sin \theta \det [\vec{\nabla}\theta \,\,\vec{\nabla}\phi] \neq 0 \quad \Leftrightarrow \quad \sin\theta \neq 0 \text{ and $\vec{\nabla}\phi$, $\vec{\nabla}\theta$ linearly independent.}
\end{align}

Finally, we point out that extending the interlayer term to depend on momentum via $t_{x,\vec{p}}=t_x + \varrho_x \vec{p}^2$ already induces a term that couples $(\tau=+,s=\downarrow)$ and $(\tau=-,s=\uparrow)$ on the level of the first term in \equref{SWTrafo}. This leads to the additional contribution to \equref{ProjectedHamZero} given by
\begin{equation}
    \Delta H = - \varrho_3 \sum_{i=x,y} \left( \{\hat{p}_i, e_{ix}(\vec{r})\}\sigma_x - \{\hat{p}_i, e_{iy}(\vec{r})\}\sigma_y \right),
\end{equation}
which is identical to \equref{SOCTermInduced} except for the different prefactor.

\section{Lensing}
In this section, we provide more details on the derivation of the lensing effects of an incoming semiclassical ray from the potential term. 
\\
\\
As a starting point, we define a circular domain wall texture as 
\begin{equation}
\theta(\boldsymbol{r})=\frac{\pi}{2},
\qquad
\phi(\boldsymbol{r})=\phi(r)=\frac{\pi}{2}\tanh\!\left(\frac{r-R}{w}\right),
\qquad
\vec n(r,\varphi)=(\cos\phi(r),\sin\phi(r),0).
\end{equation}
where $w$ is the domain wall width and $R$ is its radius. We note that, 
\begin{equation}
\vec{n}
\cdot(\partial_x\vec{n}\times\partial_y\vec{n})
=
0 
\end{equation}
for this texture and, as a result, the orbital magnetic field contribution vanishes.
\\
\\
We want to evaluate the potentials, $V_{z}$ and $V_{0}$, for this domain wall texture. For this purpose, we first need to compute the metric, 
\begin{equation}
g_{ij}=\frac{1}{4}(\partial_i\phi)(\partial_j\phi) 
\quad \text{so that} \quad
g_{xx}=\frac{1}{4}\phi'(r)^2\cos^2\varphi,
\quad \text{and} \quad
g_{yy}=\frac{1}{4}\phi'(r)^2\sin^2\varphi.
\end{equation}
From this form of the metric, we find that, 
\begin{equation}
\begin{split}
V_z(\boldsymbol{r})
&=
\frac{\hbar^2}{4}\eta^{ij}g_{ij}
=
\frac{\hbar^2}{4}(g_{xx}-g_{yy})
=
\frac{\hbar^2}{4}
\phi'(r)^2
\cos2\varphi
=
f(r)
\cos2\varphi
\quad
\text{with}
\quad
f(r)=\frac{\hbar^2}{4}
\phi'(r)^2
\\
V_0(\boldsymbol{r})
&=
\frac{\hbar^2}{4}\delta^{ij}g_{ij}
=
\frac{\hbar^2}{4}(g_{xx}+g_{yy})
=
\frac{\hbar^2}{4}
\phi'(r)^2
\end{split}
\end{equation}
As a next step, we write down the semiclassical equations of motion,
\begin{equation}
\dot{\boldsymbol{r}}
=
\frac{1}{\hbar}
\nabla_{\boldsymbol{k}}
\varepsilon_\pm,
\quad \text{and} \quad
\hbar\dot{\boldsymbol{k}}
=-\nabla_{\boldsymbol{r}}\varepsilon_\pm .
\end{equation}
where we have defined the energies, 
\begin{equation}
\varepsilon_\pm(\boldsymbol{k},\boldsymbol{r})
=
\varepsilon_0(\boldsymbol k)
+\varrho_0V_0(\boldsymbol{r})
\pm \varrho_zV_z(\boldsymbol{r}).
\end{equation}
We will primarily be interested in the equation of motion,
\begin{equation}
\hbar\dot{\boldsymbol{k}}
=
-\hat{\boldsymbol{r}}
[\varrho_0 f'(r)\pm \varrho_z f'(r)\cos2\varphi]
\pm 
\hat{\boldsymbol\varphi}
\varrho_z\frac{2f(r)}{r}\sin2\varphi    
\end{equation}
to understand the change in momentum of an incoming ray due to the domain wall. Here, we have
defined the unit vectors in the radial and angular directions 
$\hat{\boldsymbol{r}}=(\cos\varphi,\sin\varphi)$ and
$\hat{\boldsymbol\varphi}=(-\sin\varphi,\cos\varphi)$. Moreover, we have used the polar representation of the gradient,
$\nabla_{\boldsymbol{r}}=\hat{\boldsymbol{r}}\partial_r
+\hat{\boldsymbol\varphi}\frac{1}{r}\partial_\varphi$. Lastly, it will be useful to separately write down the components of this equation in the radial and angular direction, 
\begin{equation}
\hbar\dot k_r
=
-[\varrho_0 f'(r)\pm \varrho_z f'(r)\cos2\varphi]
\quad
\text{and}
\quad
\hbar\dot k_\varphi
=
\pm \varrho_z\frac{2f(r)}{r}\sin2\varphi 
\end{equation}
\\
\\
Let us now assume that we have some incoming ray with velocity,
\begin{equation}
\boldsymbol v_{\text{in}}
\approx
\frac{1}{\hbar}\nabla_{\boldsymbol{k}}\varepsilon_0(\boldsymbol{k})
\big|_{\boldsymbol{k}=\boldsymbol{k}_{\text{in}}} .
\end{equation}
The ray hits the domain wall at, 
\begin{equation}
\boldsymbol{r}_0=R \hat{\boldsymbol{r}}(\varphi_0),
\end{equation}
which defines an angle, $\varphi_0$. The velocity components at this point of impact are,
\begin{equation}
v_r
=
\boldsymbol v_{\text{in}}\cdot\hat{\boldsymbol{r}}(\varphi_0)
\quad
\text{and}
\quad
v_\varphi=\boldsymbol v_{\text{in}}\cdot\hat{\boldsymbol\varphi}(\varphi_0).
\end{equation}
We assume now that the domain wall, i.e. the region where the potentials $V_0$ and $V_z$ are finite, is very thin, $wr\ll R$. In this case, we will assume that $\varphi(t)$ and $r(t)$ are unchanged for the moment that the ray transverses through the domain wall,
\begin{equation}
\varphi(t)
\approx
\varphi_0,
\quad
r(t)\approx R,
\quad
dt\approx\frac{dr}{v_r}.
\end{equation}
We then first compute the change in the angular component of the momentum by integrating the semiclassical equation in the angular direction,
\begin{equation}
\begin{split}
\Delta k_{\varphi,\pm}
&=
\pm\frac{\varrho_z}{\hbar}
\int dt
\frac{2f(r(t))}{r(t)}
\sin 2\varphi(t)
\\
&\approx
\pm\frac{\varrho_z}{\hbar}
\frac{2\sin 2\varphi_0}{R}
\frac{1}{v_r}
\int_{-\infty}^{\infty} dr f(r)
\\
&=
\pm\frac{\varrho_z}{\hbar}
\frac{2\sin 2\varphi_0}{R}
\frac{1}{v_r}
\left(\frac{\hbar^2\pi^2}{12w}\right)
\\
&=
\pm\frac{\hbar \varrho_z \pi^2}{6 w R}
\frac{\sin 2\varphi_0}{v_r}.
\end{split}
\end{equation}
\\
\\
We can perform a similar calculation for the radial component of the momentum, 
\begin{equation}
\begin{split}
\Delta k_{r,\pm}
&=
\int_{-\infty}^{\infty} dt
\dot{k}_r
=
-
\frac{1}{\hbar}
\int_{-\infty}^{\infty} dt
[\varrho_0 f'(r(t))
\pm 
\varrho_z f'(r(t)) 
\cos 2\varphi(t)]
\\
&\approx
-
\frac{1}{\hbar v_r}
[\varrho_0
\pm
\varrho_z \cos 2\varphi_0
]
\int_{-\infty}^{\infty} dr
f'(r)
\\
&=
-
\frac{1}{\hbar v_r}
[\varrho_0 
\pm
\varrho_z \cos 2\varphi_0]
[f(\infty)-f(-\infty)]
\\
&=
0,
\end{split}
\end{equation}
where in the last equality we have used that $f(\infty)=f(-\infty)\approx 0$.
\\
\\
Hence, we see that the ray experiences primarily a momentum deflection along the angular coordinate.

\section{Simulation details}
In this section, we provide technical details on our simulations for the figures of the main text. 
\subsection{Simulation details for Fig.~2}
We first summarize our approach for obtaining the results shown in Fig.~2 of the main text. 
\\
\\
In all panels, we consider a static, coplanar circular Néel domain wall with,
\begin{equation}
\theta(\boldsymbol{r})=\frac{\pi}{2},
\quad
\phi(\boldsymbol{r})=\frac{\pi}{2}\tanh\!\left(\frac{r-R}{w}\right),
\quad
\text{and}
\quad
\phi'(r)=\frac{\pi}{2w}\,\mathrm{sech}^2\!\left(\frac{r-R}{w}\right),
\end{equation}
where we are using polar coordinates, $\boldsymbol{r}=(r,\varphi)$.
Since the texture is coplanar and time-independent, 
the emergent orbital fields vanish,
$\mathcal{B}_\tau=\mathcal{E}_{i,\tau}=0$.
We then evaluate the metric tensor of the texture, 
\begin{equation}
g_{ij}(\boldsymbol{r})
=
\frac{1}{4}\,\partial_i\vec n(\boldsymbol{r})\cdot\partial_j\vec n(\boldsymbol{r})
=
\frac{\phi'(r)^2}{4}\,\hat r_i \hat r_j ,
\qquad
\hat{\boldsymbol{r}}=(\cos\varphi,\sin\varphi).
\end{equation}
with $K^{ij}_\tau=\varrho_0\delta^{ij}+\tau\varrho_z\eta^{ij}$. 
\\
\\
For the $d$-wave altermagnetic texture in \textit{panel (a)}, we plot, 
\begin{equation}
V_z(\boldsymbol r)
=
\hbar^2
\eta^{ij}g_{ij}(\boldsymbol r)
=
\hbar^2[g_{xx}(\boldsymbol r)-g_{yy}(\boldsymbol r)]
=
(\hbar^2/4)\phi'(r)^2\cos(2\varphi).
\end{equation}
For the $g$-wave altermagnetic texture in \textit{panel (b)}, we plot,
\begin{equation}
V_z^{(g)}(\boldsymbol r)
=
\hbar^4
T^{ijkl}g_{ij}(\boldsymbol r)
g_{kl}(\boldsymbol r)
=
\hbar^4
g_{xy}(g_{xx}-g_{yy})
=
(\hbar^4/64)\phi'(r)^4\sin(4\varphi).
\end{equation}
The parameters used in the simulations are given in Table~\ref{table_fig2}. We set $\hbar=1$.
\begin{table}[h!]
\centering
\renewcommand{\arraystretch}{1.15}
\begin{tabular}{@{}ccc@{}}
$R$ & $w$   \\
\midrule
$4.0$ & $0.55$  \\
\bottomrule
\end{tabular}
\caption{\textbf{Simulation parameters for Fig.~2.}
}
\label{table_fig2}
\end{table}

\subsection{Simulation details for Fig.~3}
We next summarize our approach for obtaining the results shown in Fig.~3 of the main text. 
\\
\\
In all panels, we consider again a static, coplanar circular Néel domain wall with the same parameterization as in Fig.~3. The effective metric is given by, 
\begin{equation}
g^{ij}_{\mathrm{eff},\tau}(\boldsymbol{r})
=
K^{ij}_\tau
-
\frac{2\hbar^2}{J}
\bigl[K_\tau\, g(\boldsymbol{r}) K_\tau\bigr]^{ij},
\end{equation}
with
$K^{ij}_\tau=\varrho_0\delta^{ij}+\tau\varrho_z\eta^{ij}$ 
\\
\\
For \textit{panels (a) and (b)}, we compute the eigenvalues $\lambda_{\max}(\boldsymbol{r})\geq \lambda_{\min}(\boldsymbol{r})$ 
of $g^{ij}_{\mathrm{eff},\tau}(\boldsymbol{r})$ and plot the anisotropy, 
\begin{equation}
\Delta\lambda(\boldsymbol{r})=\lambda_{\max}(\boldsymbol{r})-\lambda_{\min}(\boldsymbol{r}),    
\end{equation}
which we normalize by its maximum value across both $\tau$ sectors. 
\\
\\
For \textit{panels (d) and (d)}, we evaluate the eigenvector associated with $\lambda_{\max}$ along the domain-wall circle $r=R$.
We then sample angles $\varphi$ on the circle and draw short line segments in the direction of that eigenvector to show the local principal axis orientation. The parameters used in our simulations of Fig.~3 are shown in Table~\ref{table_fig3}, where we have again set $\hbar=1$. 

\begin{table}[h!]
\centering
\renewcommand{\arraystretch}{1.15}
\begin{tabular}{@{}ccccc@{}}
$\varrho_0$ & $\varrho_z$ & $J$  & $R$ & $w$   \\
\midrule
$1.0$ & $0.30$ & $10.0$  & $4.0$ & $0.55$   \\
\bottomrule
\end{tabular}
\caption{
\textbf{Simulation parameters for Fig.\,3.}}
\label{table_fig3}
\end{table}

\subsection{Simulation details for Fig.~4}
We next summarize our approach for obtaining the results shown in Fig.~4 of the main text.
\\
\\
In all panels, we consider again a static, coplanar circular Néel domain wall with the same parameterization as in Fig.~2 and Fig.~3. 
The semiclassical dispersion for each $\tau=\pm$ sector is,
\begin{equation}
\varepsilon_\tau(\boldsymbol{p},\boldsymbol{r})
=
\boldsymbol{p}^{T}\,
g^{\mathrm{eff}}_{\tau}(\boldsymbol{r})
\boldsymbol{p}
+
V_\tau(\boldsymbol{r})
\quad
\text{with}
\quad
V_\tau(\boldsymbol{r})
=
\varrho_0 V_0(\boldsymbol{r})
+
\tau \varrho_z V_z(\boldsymbol{r}).
\end{equation}
We obtain the ray trajectories from Hamilton's equations,
\begin{equation}
\dot r_i=\frac{\partial \varepsilon_\tau}{\partial {p}_i},
\qquad
\dot{p}_i=-\frac{\partial \varepsilon_\tau}{\partial r_i}.
\end{equation}
For the simulations, we again set $\hbar=1$ and use the initial conditions
$\boldsymbol{r}_{\text{in}}=(-x_0,b)$ and $\boldsymbol{p}_{\text{in}}=(p_0,0)$,
with several impact parameters $b\in[-2.2,2.2]$. While we fix the incoming momentum $p_0$ to be the same for both $\tau=\pm 1$, we remark that the group velocity along the $x$-direction, 
$v_\tau 
=
\frac{\partial \varepsilon_\tau}{\partial p_x}
= 2(\varrho_0 + \tau \varrho_z) p_0$,
will still be $\tau$-dependent. 
With our simulation parameters, $\varrho_0=1$, $\varrho_z=0.35$, we obtain $v_{+}/v_{-}\approx 2.08$. 
So the $\tau=+$ rays are more than twice as fast as the $\tau=-$ rays. Because the group velocity is larger for $\tau=+$, the corresponding potential-induced deflection is suppressed.
\\
\\
The parameters used in Fig.~4 are summarized in Table~\ref{table_fig4}.

\begin{table}[h!]
\centering
\renewcommand{\arraystretch}{1.15}
\begin{tabular}{@{}lcccccc@{}}
$ $ & $\varrho_0$ & $\varrho_z$ & $w$ & $J$ & $\Pi_0$ & $R$ \\
\midrule
Fig.~4(a,c)    & $1.0$ & $0.35$ & $0.50$ & $5.5$  & $3.0$ & $4.0$ \\
Fig.~4(b,d) & $1.0$ & $0.35$ & $0.60$ & $30.0$ & $2.0$ & $4.0$ \\
\bottomrule
\end{tabular}
\caption{
\textbf{Simulation parameters for Fig.\,4.}
}
\label{table_fig4}
\end{table}

\subsection{Simulation details for Figure 5 of the main text}
We lastly summarize our approach for obtaining the results shown in Fig.~5 of the main text. 
\\
\\
In \textit{panel (a)}, we show a circular Néel domain wall with the same parameterization as in Figs.~2-4.
\\
\\
In \textit{panel (b)}, we obtain the  constant-energy contour from the local dispersion, $E_\pm(\boldsymbol p;\boldsymbol{r})=\mu$, with,
\begin{equation}
E_\pm(\boldsymbol p;\boldsymbol{r})
=
\varepsilon_{0,\boldsymbol p}
+
\varrho_0 V_0(\boldsymbol{r})
\pm
\sqrt{
(t_{z,\boldsymbol p}+\varrho_z V_z(\boldsymbol{r}))^2
+
\vec{\mathtt g}_{\boldsymbol p}^{\,2}(\boldsymbol{r})
},
\end{equation}
evaluated at the marked point, $(r,\varphi)=(R,\varphi_0)$. Here, $t_{z,\boldsymbol p}=\varrho_z(p_x^2-p_y^2)$ and the spin-orbit field is given by, 
$
\vec{\mathtt g}_{\boldsymbol p}(\boldsymbol{r})
=
g_0
(
\sin\theta(\boldsymbol{r})
\boldsymbol p\cdot\nabla\phi(\boldsymbol{r}),
\boldsymbol p\cdot\nabla\theta(\boldsymbol{r})
)^T
$
\\
\\
In \textit{panel (c)}, we plot a skyrmion texture defined by
\begin{equation}
\theta(r)=2\arctan\left(\frac{a^2}{r^2}\right)
\quad,
\quad
\phi(\boldsymbol{r})=\varphi
\end{equation}
In \textit{panel (d)}, the constant-energy contours are computed as in panel (b), 
evaluated at the marked point
$
(r,\varphi)=(r_0,\varphi_0).
$
\\
\\
The parameters used in our simulations are given in Table~\ref{table_fig5}.
\begin{table}[h!]
\centering
\renewcommand{\arraystretch}{1.15}
\begin{tabular}{@{}lccccccccc@{}}
$\,$ & $\varrho_0$ & $\varrho_z$ & $g_0$ & $\mu$ & $(r_0,\varphi_0)$ & $R$ & $w$ & $a$ \\
\midrule
Fig.~5(a,b) & $1.0$ & $0.3$ & $-0.4$ & $1.0$ & $(R,\pi/4)$ & $3.6$ & $0.7$ & -- \\
Fig.~5(c,d) & $1.0$ & $0.3$ & $-0.4$ & $1.0$ & $(1.0,\pi/4)$ & -- & -- & $2.0$ \\
\bottomrule
\end{tabular}
\caption{
\textbf{Simulation parameters for Fig.\,5.}
}
\label{table_fig5}
\end{table}

\end{document}